\documentclass[amsmath,twocolumn,pra,longbibliography,superscriptaddress]{revtex4-2} 
\usepackage{amsthm,amssymb,amsfonts,graphicx,verbatim, xcolor,bm,ulem} 
\usepackage[hidelinks]{hyperref} 
\usepackage[utf8]{inputenc} 
\usepackage{siunitx} 
\usepackage{cleveref}
\usepackage{multirow}
\usepackage{colortbl}
\usepackage{threeparttable,booktabs}

\renewcommand{\vec}[1]{{\bm #1}}
\renewcommand{\Im}{{\rm Im}\,}
\renewcommand{\Re}{{\rm Re}\,}

\newcommand{\ket}[1]{|#1\rangle}

\newcommand{\braket}[2]{\langle #1|#2\rangle}
\newcommand{\braOket}[3]{\langle #1|#2|#3\rangle}
\newcommand{\mr}{moir\'e~}
\newcommand{\Mr}{Moir\'e~}

\usepackage{tikz}
\usepackage{pgfplots}
\usepgfplotslibrary{groupplots,dateplot}
\usetikzlibrary{patterns,shapes.arrows,math,calc,arrows.meta}
\pgfplotsset{compat=newest}

\begin{document}

\title{Chiral model of twisted bilayer graphene realized in a monolayer
}
\author{Valentin Cr\'epel}
\affiliation{Center for Computational Quantum Physics, Flatiron Institute, New York, New York 10010, USA}
\author{Aaron Dunbrack}
\affiliation{Department of Physics and Astronomy, Stony Brook University, Stony Brook, New York 11794, USA}
\author{Daniele Guerci}
\affiliation{Center for Computational Quantum Physics, Flatiron Institute, New York, New York 10010, USA}
\author{John Bonini}
\affiliation{Center for Computational Quantum Physics, Flatiron Institute, New York, New York 10010, USA}
\author{Jennifer Cano}
\affiliation{Center for Computational Quantum Physics, Flatiron Institute, New York, New York 10010, USA}
\affiliation{Department of Physics and Astronomy, Stony Brook University, Stony Brook, New York 11794, USA}

\begin{abstract}
We demonstrate that a single layer of graphene subject to a superlattice potential nearly commensurate to a $\sqrt{3} \times \sqrt{3}$ supercell exactly maps to the chiral model of twisted bilayer graphene, albeit with half as many degrees of freedom. 
We comprehensively review the properties of this ``half-chiral model,'' including the interacting phases stabilized at integer fillings and the effects of substrate-induced symmetry breaking. We list candidate substrates that could produce a superlattice potential on graphene with the correct periodicity to access the flat band limit. Experimental measurements on a half-chiral moir\'e heterostructure, in which valley-skyrmions cannot form, could yield insights on the physics they mediate in twisted bilayer graphene.
\end{abstract}

\maketitle

\section{Introduction}

In the five years following its groundbreaking discovery~\cite{cao2018correlated,cao2018unconventional}, twisted bilayer graphene (TBG) has realized many of the quantum phases thus far discovered in condensed matter physics~\cite{andrei2020graphene,andrei2021marvels}.
Since then, developing an understanding of TBG has been a major area of research, triggering a colossal body of theoretical work aiming at disentangling the physics of this and other moir\'e materials~\cite{tarnopolsky2019origin,dos2007graphene,bistritzer2011moire,bernevig2021twistedI,song2021twisted,bernevig2021twistedIII,lian2021twisted,bernevig2021twisted,xie2021twisted,crepel2022anomalous,mao2023supermoire,dunbrack2021magic,crepel2022unconventional,ghorashi2022topological,dunbrack2023intrinsically,crepel2023topological}.

In this quest, a particularly simple yet powerful and predictive approach was the introduction of an analytically tractable model --- dubbed the chiral model~\cite{tarnopolsky2019origin} --- 
obtained by slight simplification from earlier continuum models of TBG~\cite{dos2007graphene,bistritzer2011moire,2023arXiv230510477C}.
The chiral model captures the essence of correlation in TBG in that it features eight-fold degenerate \textit{exact} flat bands for certain magic values of the twist angle~\cite{bernevig2021twistedI}. Mathematically, the perfect degeneracy of these bands stems from an emergent chiral symmetry which is only weakly broken in TBG~\cite{tarnopolsky2019origin,bernevig2021twistedIII}, making the chiral model a good qualitative guide to appreciate the physics of TBG while making analytical progress. 
For instance, the chiral symmetry provides an intuitive explanation for the insulating states at integer filling by analogy with quantum Hall ferromagnetism, which helps rationalize the quantum anomalous Hall (QAH) state at filling $n=3$ in TBG~\cite{Sharpe_2019,Serlin_2020,Nuckolls_2020,Saito_2021,Wu_2021}. The analogy to Hall systems also hints at the potential role of skyrmions as bosonic glue in TBG's superconducting state~\cite{garnier2019topological,khalaf2021charged,chatterjee2022skyrmion}. Finally, the chiral flat bands can be formally written as elements of a Landau level~\cite{wang2021chiral,sheffer2021chiral,estienne2023ideal}, allowing for the definition of exact multi-component fractional Chern insulators~\cite{ledwith2020fractional,crepel2018matrix,wang2021exact,crepel2019matrix} which may describe those measured in TBG at finite  field~\cite{xie2021fractional}.

It remains an open question which aspects of the chiral model can be smoothly transferred to TBG. 
Answering this question turns out to be as computationally challenging as the direct simulation of the continuum model for TBG~\cite{bultinck2020ground,hofmann2022fermionic,xie2021twisted}. As an alternate path forward, we propound the design of novel moir\'e materials which exactly realize the chiral model.
If such a material existed, comparison of its phase diagram with that of TBG would elucidate which phases of the latter are smoothly connected to the chiral model. Even more interesting, theoretical questions concerning which physics can be directly attributed to valley-skyrmions in TBG become similarly addressable through a moir\'e analog of the chiral model with half as many degrees of freedom --- the ``half-chiral model'' --- in which such skyrmions cannot form.

To this end, we describe an exact realization of the half-chiral model from a superlattice potential nearly commensurate to a $\sqrt{3}\times\sqrt{3}$ supercell of graphene --- dubbed a near-$\sqrt{3}$ matched potential hereafter --- imposed onto a graphene monolayer. Since this potential varies on a scale comparable to graphene's lattice constant, it cannot be engineered by either twist or strain, but can be produced by electrostatic coupling to a substrate. Far from a theoretic abstraction, Ref.~\cite{lu2022dirac} presented a similar analysis motivated by their experimental data on a SiC/graphene heterostructure, noting the relation to the half-chiral model for large values of the graphene-substrate coupling. 

In the present manuscript, we provide a complete picture for the realization of the half-chiral model using a near-$\sqrt{3}$ matched substrate-induced potential. 
We introduce the model in Sec.~\ref{sec:IdealModel}.
In Sec.~\ref{Sec:InteractionsIntegerFilling} we consider the effect of interactions and show that the many body ground states at integer filling can realize QAH states similar to chiral TBG.
In Sec.~\ref{sec_symbreak}, we provide a list of candidate substrates together with their most important symmetry breaking effects. Our work paves the way for more proposals and designs of moir\'e materials capable of answering questions about the many-body physics of TBG, whose full phase diagram remains out of current numerical reach.

\section{Half-chiral model} \label{sec:IdealModel}

In this section, we model the graphene-substrate heterostructure described in the introduction, in which a substrate provides a near-$\sqrt{3}$ matched electrostatic potential that directly couples the two valleys of graphene on an emergent moir\'e scale.

We assume the potential is onsite and hence diagonal in the sublattice index.
It maps exactly to a term that was shown in Ref.~\cite{DunbrackTwistedTIs} to allow for a vanishing Dirac cone velocity.
If the moir\'e coupling furthermore respects the original rotation, mirror, and time reversal symmetries of the monolayer, the vanishing Dirac cone velocity implies the existence of perfectly flat bands~\cite{FlatBandsFromSymmetries}. A related model in which the potential is not diagonal in sublattice has been studied in Ref.~\cite{scheer2023kagome}.

\subsection{Setup}
\label{sec:setup}

The low-energy physics of a graphene monolayer derives from its $K$ and $K'$ points, described by the effective continuum Hamiltonians
\begin{equation}\label{eq_bareHKKp}
H_K(\bar k)= - v \bar k\times\tau , \quad H_{K'}(\bar k)= v \bar k\times\tau^* ,
\end{equation}
with $\bar k$ the deviation from the $K$/$K'$ points, $\vec{\tau}$ the sublattice Pauli matrices, and $v$ the Fermi velocity of graphene. Here, we use the convention $e_\pm =  (3/2, \pm \sqrt{3}/2)$ for the Bravais lattice vectors, $r_A = (1,0)$ and $r_{B} = (1/2, \sqrt{3}/2)$ for the position of orbitals in the unit cell, and have set graphene's lattice constant as the unit of length. This choice centers the two valleys around the $\tau_z K_{n=0,1,2}$ corners of the Brillouin zone (BZ), where
\begin{equation}
K_n = K \left( \sin \frac{2n\pi}{3}, \cos \frac{2n\pi}{3} \right) , \quad K = \frac{4\pi}{3\sqrt{3}} .
\end{equation}

\begin{figure}
\centering
\includegraphics[width=\columnwidth]{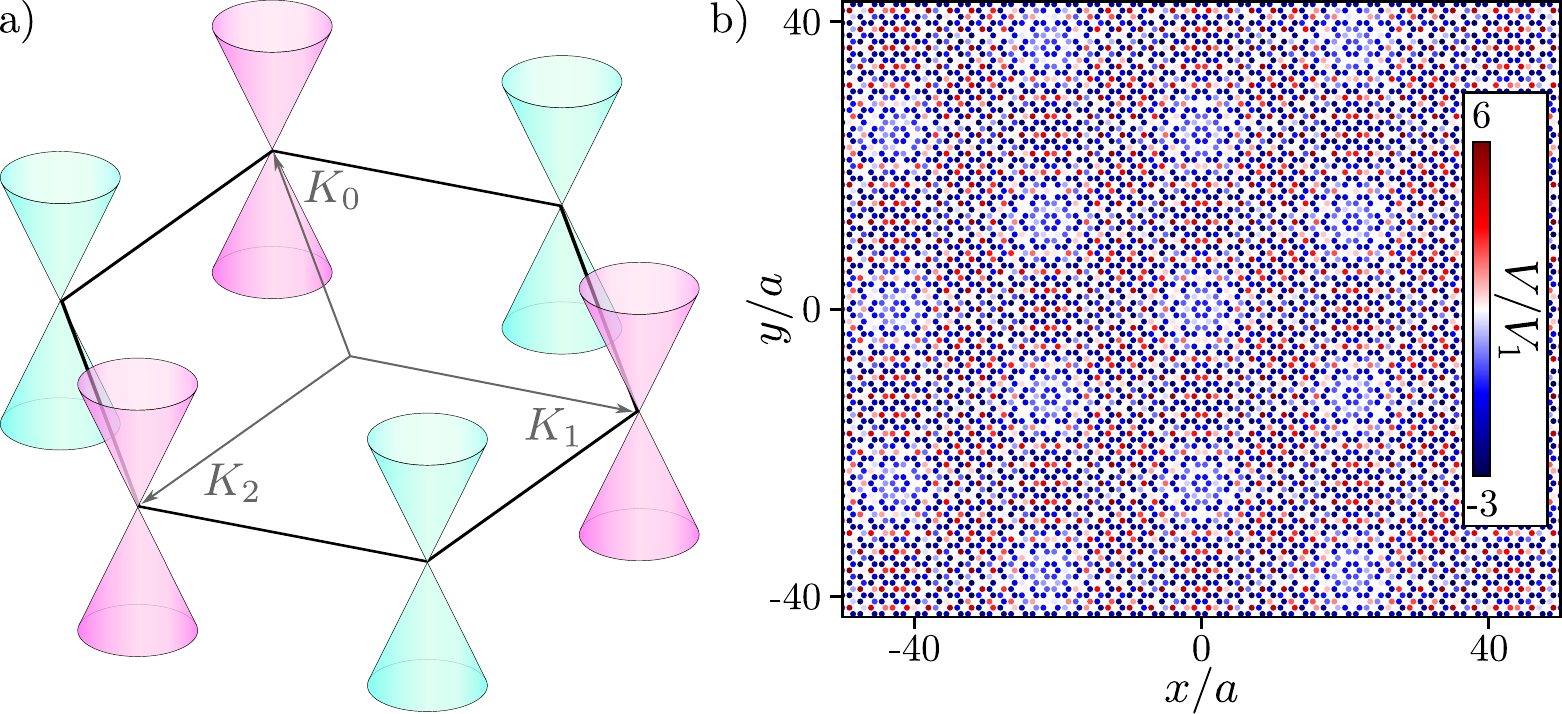}
\caption{a) The two valleys of graphene are coupled by a near-$\sqrt{3}$ matched potential with reciprocal lattice vectors $(1+\epsilon) K_{0,1,2}$, which forms a moir\'e pattern when evaluated on graphene sites, as shown in (b) for $\epsilon = 0.07$.
}
\label{fig:ModelAndMoire}
\end{figure}

The two valleys are coupled by a near-$\sqrt{3}$ matched onsite potential of the form 
\begin{equation} \label{eq_nearroot3matched}
V(r) = 2 V_1 \sum_{n=0}^2 \cos [ ( K_n + q_n ) \cdot r] , \quad q_n = \epsilon K_n
\end{equation}
with $|\epsilon| \ll 1$ (see Fig.~\ref{fig:ModelAndMoire}a). For $\epsilon=0$, the potential folds the $\pm K_n$ onto the $\Gamma$ point of graphene's BZ and enlarges the unit cell to $\sqrt{3}\times \sqrt{3}$. However, in this case, the form of Eq.~\eqref{eq_nearroot3matched} shows that $V(r)$ identically vanishes on all graphene sites. For nonzero but small $\epsilon$, this potential forms a moir\'e pattern with lattice constant $a_m = \epsilon^{-1}$ due to the small mismatch $q_n$ between the reciprocal lattice vectors of the potential and the edges of the graphene Brillouin zone~\cite{dunbrack2023intrinsicallymultilayer}, as shown in Fig.~\ref{fig:ModelAndMoire}b.

The effective Hamiltonian of this moir\'e system is obtained by projecting $V(r)$ near the $K/K'$ points, yielding
\begin{equation} \label{eq_originalcontinuummodel}
H^{0}=\begin{bmatrix}
H_K&\tilde V(r)\\
\tilde V(r)^\dagger&H_{K'} 
\end{bmatrix} , \quad \tilde V(r)=V_1\sum_{n=0}^2 e^{i (q_n \cdot r + n \phi \tau_z)} ,
\end{equation}
where $\phi = 2\pi/3$. The phases in $\tilde V$ are set by $K_n \cdot (r_A - r_B) = 2 n \phi$ $[{\rm mod} \, 2\pi]$ (see Appendix~\ref{Apx:basics} for details). 
This model preserves the crystalline and time-reversal symmetries of graphene. Reintroducing the spin, implicit in Eq.~\eqref{eq_originalcontinuummodel}, $H^{0}$ is also invariant under SU(2) spin rotations, implemented by the $\vec{\sigma}$ Pauli matrices, and under products such as $\Lambda_x = i\sigma_x C_{2z}T$, which can be understood as a spinless version of $C_{2z}T$.
These symmetries are summarized in Table~\ref{tab:Symmetries}, which also lists a more subtle anticommuting symmetry, $\Lambda_z$, whose properties and consequences will be studied shortly.

\begin{table}[]
\centering
\begin{tabular}{|c||c|c|c||c|c|c|c|}
\hline
Symmetry& Original &TBG$_K$&Chiral&$\bar k_x$&$\bar k_y$&$r_x$&$r_y$  \\
\hline\hline
$C_{2x}$&$i\sigma_x\mu_x$&$i\tau_x\sigma_x\mu_x$&$i\tau_x\sigma_x\mu_x$&$+$&$-$&$+$&$-$ \\
\hline
$C_{2y}$&$i\tau_x\sigma_y$&$-i\tau_y\sigma_y\mu_z$&$-i\tau_z\sigma_y\mu_y$&$-$&$+$&$-$&$+$ \\
\hline
$C_{2z}$&$i\tau_x\sigma_z\mu_x$&$-i\tau_z\sigma_z\mu_y$&$-i\tau_y\sigma_z\mu_z$&$-$&$-$&$-$&$-$ \\
\hline
$I$&$\tau_x\mu_x$&$-\tau_z\mu_y$
&$-\tau_y\mu_z$&$-$&$-$&$-$&$-$ \\
\hline
$T$&$i\sigma_y\mu_xK$&$\tau_y\sigma_y\mu_yK$&$\tau_y\sigma_y\mu_yK$&$-$&$-$&$+$&$+$ \\
\hline\hline
SU(2)$_\text{Spin}$&$\sigma_{x,y,z}$&$\sigma_{x,y,z}$&$\sigma_{x,y,z}$&$+$&$+$&$+$&$+$ \\
\hline
$\Lambda_z$&$\tau_z\mu_z$&$\tau_z$&$\mu_z$&$+$&$+$&$+$&$+$ \\
\hline
$\Lambda_x \! = \! i\sigma_x C_{2z}T$&$\tau_xK$&$\tau_xK$&$\mu_xK$&$+$&$+$&$-$&$-$ \\
\hline
\end{tabular}
\caption{Symmetries of the model represented in the original [Eq.~\eqref{eq_originalcontinuummodel}], TBG$_K$ [Eq.~\eqref{eq_tbgbasis}] and chiral [Eq.~\eqref{eq_chiralmodel}] bases, with the exact symmetries of graphene (crystalline and time-reversal) given above the second double-line. The last four columns indicate whether the symmetry changes the sign of each component of momentum and position. $\Lambda_z$ anticommutes with the Hamiltonian, while all other symmetries commute with it. In the original basis, $\tau$, $\sigma$ and $\mu$ are sublattice, spin and valley Pauli matrices, respectively. Identity matrices $\tau_0$, $\sigma_0$, $\mu_0$ are left implicit.
} 
\label{tab:Symmetries}
\end{table}

\subsection{Mapping to chiral model} \label{ssec:mappingchiralmodel}

We now map the Hamiltonian in Eq.~\eqref{eq_originalcontinuummodel} onto the chiral model for the $K$ valley of TBG by a series of unitary transformations. 
We first make a gauge choice for the $A$ and $B$ orbitals of each valley with ${\mathcal U}_1= e^{i \pi \tau_z \mu_z / 4}$, where $\vec{\mu}$ are valley Pauli matrices. Then, we rotate the sublattice basis in $K'$ so that both valleys feature the same helicity using ${\mathcal U}_2 = {\rm diag} (\tau_0, \tau_y)$. The resulting Hamiltonian is
\begin{equation} \label{eq_tbgbasis}
H^{{\rm TBG}_K} = ({\mathcal U}_1 {\mathcal U}_2)^\dagger H^{0} ({\mathcal U}_1 {\mathcal U}_2) = \begin{bmatrix}
v \bar k\cdot\tau & T(r)\\ T(r)^\dagger & v\bar k\cdot\tau
\end{bmatrix} ,
\end{equation}
with 
\begin{equation} \label{eq_TBGbasisTunneling}
T(r)=V_1\sum_n e^{i q_n \cdot r} \left[ (\hat q_n \times \tau)\cdot \hat z \right] ,
\end{equation}
with $\hat q_n$ the direction of $q_n$. 
This is precisely the form obtained in the $K$ valley of twisted bilayer graphene in absence of $AA$ and $BB$ couplings~\cite{dos2007graphene,bistritzer2011moire}, hence the superscript TBG$_K$.

All terms in Eq.~\eqref{eq_tbgbasis} are purely off diagonal in the sublattice index. To make this block structure more explicit, we introduce one last unitary transformation
\begin{equation}
H^{\rm chiral} = {\mathcal U}_3^\dagger H^{{\rm TBG}_K} {\mathcal U}_3 , \quad 
{\mathcal U}_3=\begin{bmatrix} 1&0&0&0\\ 0&0&1&0\\ 0&1&0&0\\ 0&0&0&1 \end{bmatrix},
\end{equation}
which, as promised, maps our system to the chiral model for the $K$ valley of TBG~\cite{tarnopolsky2019origin} 
\begin{equation} \label{eq_chiralmodel}
H^{\rm chiral} = v |q_0| \begin{bmatrix} 0 & D\\ D^\dagger&0 \end{bmatrix} ,
\quad D=\begin{bmatrix} - 2 i \partial_z & \alpha U(r) \\ \alpha U(-r) & -2 i \partial_z \end{bmatrix} ,
\end{equation}
with $\partial_z=(\partial_x+i\partial_y)/|q_0|$ and
\begin{equation}
U(r) = \sum_n e^{i(q_n \cdot r - n\phi)} , \quad \alpha = \frac{V_1}{v |q_0|} .
\end{equation}

\begin{figure}
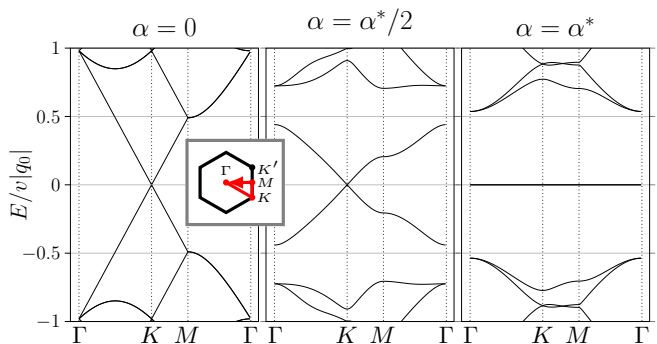

\centering
\begin{tikzpicture}[scale=0.5]
\input{SpectrumSqrt3ChiralZero}
\begin{scope}[shift={(5.2,0)}]
\input{SpectrumSqrt3ChiralHalf}
\end{scope}
\begin{scope}[shift={(10.4,0)}]
\input{SpectrumSqrt3ChiralFlat}
\end{scope}
\begin{scope}[shift={(3.1,2.5)},scale=0.4]
\input{SpectrumPath}
\end{scope}
\end{tikzpicture}
\caption{Band structure of the continuum model 
Eq.~\eqref{eq_originalcontinuummodel} plotted along the high-symmetry path of the moir\'e mini-Brillouin zone illustrated in the inset. The bandwidth of the band closest to charge neutrality exactly vanishes when the coupling strength $\alpha = V_1 / |v q_0|$ reaches the value $\alpha^* \simeq 0.586$. 
}
\label{fig:BandStructure}
\end{figure}

For convenience, we summarize the action of the successive unitary transformations on the original basis
\begin{equation} 
({\mathcal U}_1 {\mathcal U}_2 {\mathcal U}_3)^\dagger \begin{bmatrix}(A,K)\\(B,K)\\(A,K')\\(B,K')\end{bmatrix} =
e^{-i\pi/4} \begin{bmatrix}(A,K)\\-i(B,K')\\i(B,K)\\-(A,K')\end{bmatrix}.
\end{equation}
Up to an orbital-valley dependent gauge choice, this transformation sorts the lattice degrees of freedom with respect to their chirality, \textit{i.e.}, the periodic part of the Bloch orbitals projected on $A$ near $K$ and $B$ near $K'$ exhibit the same phase winding around the origin, which is opposite to the winding for $(B,K)$ and $(A,K')$~\cite{fuchs2013dirac,crepel2021universal}. These two sectors are distinguished by the chiral operator, which takes the form $\Lambda_z=\tau_z\mu_z$ in the original basis (see Table~\ref{tab:Symmetries}).

\subsection{Flat bands and degeneracy}

In the chiral model of twisted bilayer graphene, perfectly flat bands arise when the twist angle and interlayer hopping strength are tuned so that $\alpha$ -- the only dimensionless parameter of the theory -- belongs to a particular set of values ($0.586$, $2.221$, $3.75$, $\dots$)~\cite{tarnopolsky2019origin}. The direct mapping to Eq.~\eqref{eq_chiralmodel} ensures that our model also enjoys perfectly flat bands when the ratio $\alpha = V_1 / (v |q_0|)$ takes values in that set. 
In the following we employ $\alpha^*$ to denote the lowest of these magic values. Equivalently, $V^*_1$ is the magic amplitude of the potential for a given wavevector $|q_0|$. 
As an illustration, we plot in Fig.~\ref{fig:BandStructure} the band structure of the the moir\'e Hamiltonian in Eq.~\eqref{eq_originalcontinuummodel} for different values of $\alpha$ between 0 and the first magic value $\alpha^* \simeq 0.586$.

Fig.~\ref{fig:BandStructure} shows that our model features two spin-degenerate flat bands at $\alpha = \alpha^*$. One of these flat bands is annihilated by $D$ and lives in the first chirality block $\{AK, BK'\}$, while the other -- its time-reversed and particle-hole partner -- is an exact zero mode of $D^\dagger$ and lives in the opposite chirality block $\{BK, AK'\}$. We label them by their corresponding chiral operator eigenvalue $\Lambda_z$, \textit{i.e.} by $+$ and $-$, respectively. Finally, the flat bands inherit a non-zero Chern number $C(\Lambda_z) = \Lambda_z$ from the winding of their constituting orbitals~\cite{liu2019pseudo,becker2022mathematics}. Since the bands in opposite chirality sectors are time-reversal partners, they must have opposite Chern number.

\begin{figure}
\centering
\includegraphics[width=0.6\columnwidth]{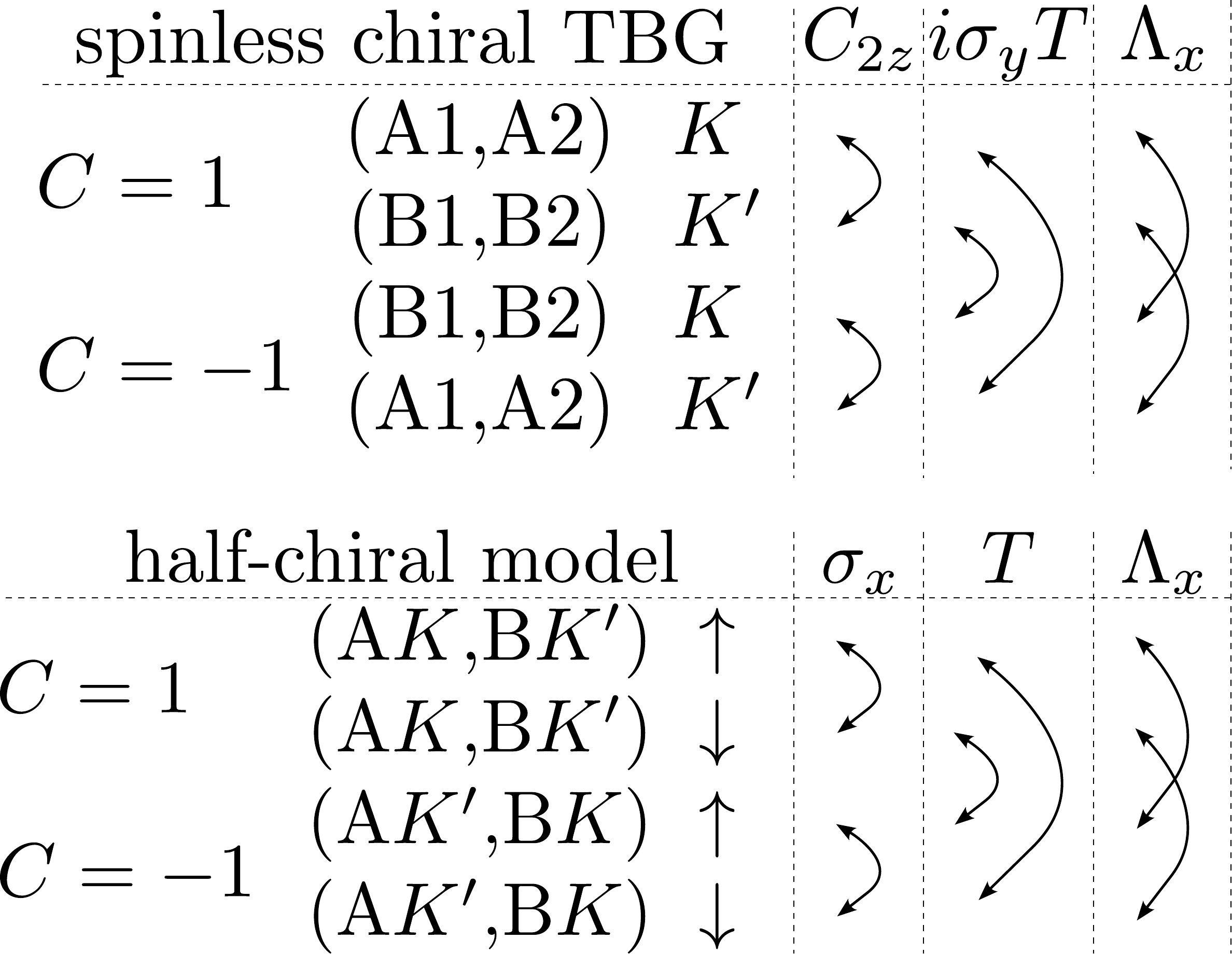}
\caption{Flat bands in the spinless chiral model of TBG (top)~\cite{tarnopolsky2019origin,bultinck2020ground} and in the half-chiral model (bottom), gathered by Chern number, or chirality. In the former, $Pn$ indicates the $P=A/B$ orbital in layer $n$. Symmetries relating the different bands are depicted on the right, showing how spin in our model plays a role equivalent to valley in spinless chiral TBG.
}
\label{fig:DiffTBG}
\end{figure}

An important distinction between our model and the chiral model of TBG is the number of flat bands. In the chiral limit of TBG, there are $2^3=8$ flat bands, labelled by spin, valley, and sublattice. In contrast, our system only displays $2^2=4$ flat bands at the magic point, labelled by spin and sublattice-valley chirality. Thus, we have provided a realization of the half-chiral model. In spite of this lower degeneracy, we can draw some analogy between the half-chiral model and the spinless model for TBG by looking at how symmetries of the two models couple the different flat bands, as summarized in Fig.~\ref{fig:DiffTBG}. There, we observe that spin in our model plays a role analogous to the valley index in spinless chiral TBG.

\section{Interactions at integer filling} \label{Sec:InteractionsIntegerFilling}

We now consider the effect of symmetry-preserving interaction terms on the Hamiltonian in Eq.~\eqref{eq_originalcontinuummodel} in the chiral limit. In Sec.~\ref{ssec_symfornfactors}, we show how symmetry constrains the form factors of the interaction. Using these considerations, in Sec.~\ref{ssec_integerfilled} we derive the many body ground states at integer filling. The discussion proceeds similarly to chiral TBG, accounting for the lower internal degeneracy of our model. Finally, we highlight the robustness of the Chern insulators obtained when an odd number of the flat bands are filled.

\subsection{Form factors} \label{ssec_symfornfactors}

Let us first show that the form factors of the interaction are diagonal in the $\sigma_z$ and $\Lambda_z$ indices using the SU(2) spin rotation and chiral symmetry of our model. We further relate these diagonal form factors using the $\sigma_x$ and $\Lambda_x$ symmetries.

Consider an inter-particle repulsive potential with Fourier transform $U_q>0$. After projection to the flat band subspace, electronic interactions take the form
\begin{equation}
H_I = \int {\rm d}q \, U_q \rho_q \rho_{-q} , \; \rho_q = \sum_{k, \alpha, \beta} c_{\alpha,k}^\dagger \braket{\psi_{k,\alpha}}{\psi_{k+q,\beta}} c_{\beta,k+q} ,
\label{eq:HI}
\end{equation}
where $\alpha = (\sigma, \gamma)$ labels the bands using their spin and chirality indices, while $c_{\alpha,k}$ and $\ket{\psi_{k,\alpha}}$ respectively correspond to the fermionic operator and the periodic Bloch wavefunction of the electronic mode in band $\alpha$ with momentum $k$.

Since each band has a well-defined chirality and spin, the form factors $\braket{\psi_{k,\alpha}}{\psi_{k+q,\beta}}$ appearing in the projected density $\rho_{q}$ drastically simplify. Specifically, using the identities $\Lambda_z^2= \sigma_z^2 = 1$, we obtain
\begin{align} 
&\braket{\psi_{(\sigma',\gamma')}}{\psi_{(\sigma,\gamma)}} \\ 
& \quad = \braOket{\psi_{(\sigma',\gamma')}}{\Lambda_z^2}{\psi_{(\sigma,\gamma)}} = \gamma \gamma'  \braket{\psi_{(\sigma',\gamma')}}{\psi_{(\sigma,\gamma)}} \notag \\
& \quad = \braOket{\psi_{(\sigma',\gamma')}}{\sigma_z^2}{\psi_{(\sigma,\gamma)}} = \sigma \sigma'  \braket{\psi_{(\sigma',\gamma')}}{\psi_{(\sigma,\gamma)}} , \notag 
\end{align}
which proves that $\braket{\psi_{(\sigma',\gamma')}}{\psi_{(\sigma,\gamma)}}=0$ when $\gamma \neq \gamma'$ or $\sigma\neq\sigma'$. The projected density operator thus assumes the diagonal form
\begin{equation} 
\rho_{q} = \sum_{k,\alpha}  F_{k,q}^{\alpha} c_{\alpha,k}^\dagger c_{\alpha, k+q} , \quad F_{k,q}^{\alpha} = \braket{\psi_{k,\alpha}}{\psi_{k+q,\alpha}} .
\end{equation}

Additional progress can be made by exploiting the full spin symmetry of the model using $\sigma_x$ and the emergent $\Lambda_x$ symmetry (see Table~\ref{tab:Symmetries}), which also both square to the identity
\begin{align}
&F_{k,q}^{(\uparrow,\gamma)} \!= \!\braOket{\psi_{(\uparrow,\gamma)}}{\sigma_x^2}{\psi_{(\uparrow,\gamma)}} \!= \!\braket{\psi_{(\downarrow,\gamma)}}{\psi_{(\downarrow,\gamma)}} \!= \!F_{k,q}^{(\downarrow,\gamma)} , \notag \\
&F_{k,q}^{(\sigma,+)} \!= \!\braOket{\psi_{(\sigma,+)}}{\Lambda_x^2}{\psi_{(\sigma,+)}} \!= \!\braket{\psi_{(\sigma,-)}^*}{\psi_{(\sigma,-)}^*} \!= \![F_{k,q}^{(\sigma,-)} ]^* ,
\end{align}
where the complex conjugation in the second line come from the anti-unitarity of $\Lambda_x$. This fixes
\begin{equation} \label{eq_symresolvedformfactor}
F_{k,q}^{(\sigma, \gamma)} = F_{k,q} e^{i \Phi_{k,q} \gamma} ,
\end{equation}
for some real functions $F_{k,q}$ and $\Phi_{k,q}$, similarly to the result obtained in the chiral model describing a single valley of TBG~\cite{bultinck2020ground}.
It follows that our half-chiral model has the same U$(2)\times$U$(2)$ symmetry as the spinless chiral model of TBG (see Fig.~\ref{fig:DiffTBG}).

Note that it is not possible to gauge away the phase $\Phi_{k,q}$ because of the non-zero Chern number of the chiral bands. Indeed, noticing that the Berry connection can be written as $\mathcal{A}_k^{(\sigma,\gamma)} = \Im (\nabla_q F_{k,q}^{(\sigma,\gamma)})_{q=0} = \gamma (\nabla_q \Phi_{k,q} )_{q=0}$, we must have 
\begin{equation}
\sum_{k} [ ( \partial_{k_x} \partial_{q_y} - \partial_{k_y} \partial_{q_x} ) \Phi_{k,q} ]_{q=0}  = 2 \pi .
\end{equation}
This ensures that $F_{k,q}^{(\sigma,+)} \neq F_{k,q}^{(\sigma',-)}$ at least at one point in the Brillouin zone. Thus, Eq.~\eqref{eq_symresolvedformfactor} cannot be further simplified.

\subsection{Integer fillings} \label{ssec_integerfilled}

Equipped with the form factors in Eq.~\eqref{eq_symresolvedformfactor}, we now investigate how interactions lift the extensive degeneracy of the flat band subspace at integer fillings using arguments borrowed from quantum Hall ferromagnetism~\cite{sondhi1993skyrmions,girvin2000spin}. Our analysis relies on two facts: ($i$) the interacting Hamiltonian is positive, \textit{i.e.} for any many body state $\ket{\Psi}$, $\braOket{\Psi}{H_I}{\Psi} = \int {\rm d}q U_q || \rho_{-q} \ket{\Psi}||^2 \geq 0$; and ($ii$) any state in which each of the chiral bands is either filled or empty is annihilated by $\rho_q$ for all $q$. 
Since the states described in ($ii$) have no interaction energy, and no kinetic energy by virtue of the flat bands, they are necessarily ground states following ($i$).
Any other state will acquire some interaction energy from one of the $\rho_{q}$. This energy, though possibly infinitesimal (\textit{e.g.} for magnons), ensures that the ground state is a band ferromagnet in the chiral basis. 

We now discuss each integer filling in turn.

\paragraph{\underline{At $n=\pm 2$},} the four bands at zero energy are either all fully filled or all empty. The total Chern number of the occupied band is zero and all spin components are equally occupied. The system is a topologically-trivial band insulator.

\paragraph{\underline{At $n=\pm 1$},} an odd number of $\alpha$ bands must be filled, following ($i$) and ($ii$) above. The ground state therefore carries a non-trivial Chern number, which manifests in an quantum anomalous Hall effect and a spontaneous spin-polarization.

\paragraph{\underline{At $n=0$},} our prescription for filling two of the four flat bands gives six degenerate ground states with sharply different physical properties. Some are paramagnetic with a total Chern number $|C_{\rm tot}|=2$, some are spin polarized with $|C_{\rm tot}|=0$, and others are paramagnetic and carry no total Chern number. To predict the behavior at charge neutrality, we exploit a hierarchy between the various symmetries of our model (see Table~\ref{tab:Symmetries}). 
The crystalline symmetries are protected by the carbon links of graphene, while the $\Lambda_z$ symmetry is an emergent symmetry, which we can realistically assume is more fragile.
Provided interaction scales remain stronger than kinetic ones, the weak breaking of $\Lambda_z$ only changes the form factors, which remain spin-diagonal but acquire non-zero off-diagonal coefficients between Bloch state of opposite chiralities.
In this situation, the many-body states with vanishing interaction energy must have all bands that are mixed by a form factor either entirely empty or fully filled. This lifts the previous six-fold degeneracy and results in a spin-polarized ground state with $|C_{\rm tot}| = 0$, where two flat bands with opposite chirality and equal spin are completely occupied.

\subsection{Chern insulator robustness}

At the magic angle and for $n=\pm 1$, we have just seen that the half-chiral model spontaneously breaks time-reversal symmetry and exhibits a quantized anomalous Hall effect.
To determine the robustness of this state we now study the spectra of charged and neutral excitations, both at and away from the magic value of $\alpha^*$.

\begin{figure}
\centering
\includegraphics[width=\columnwidth]{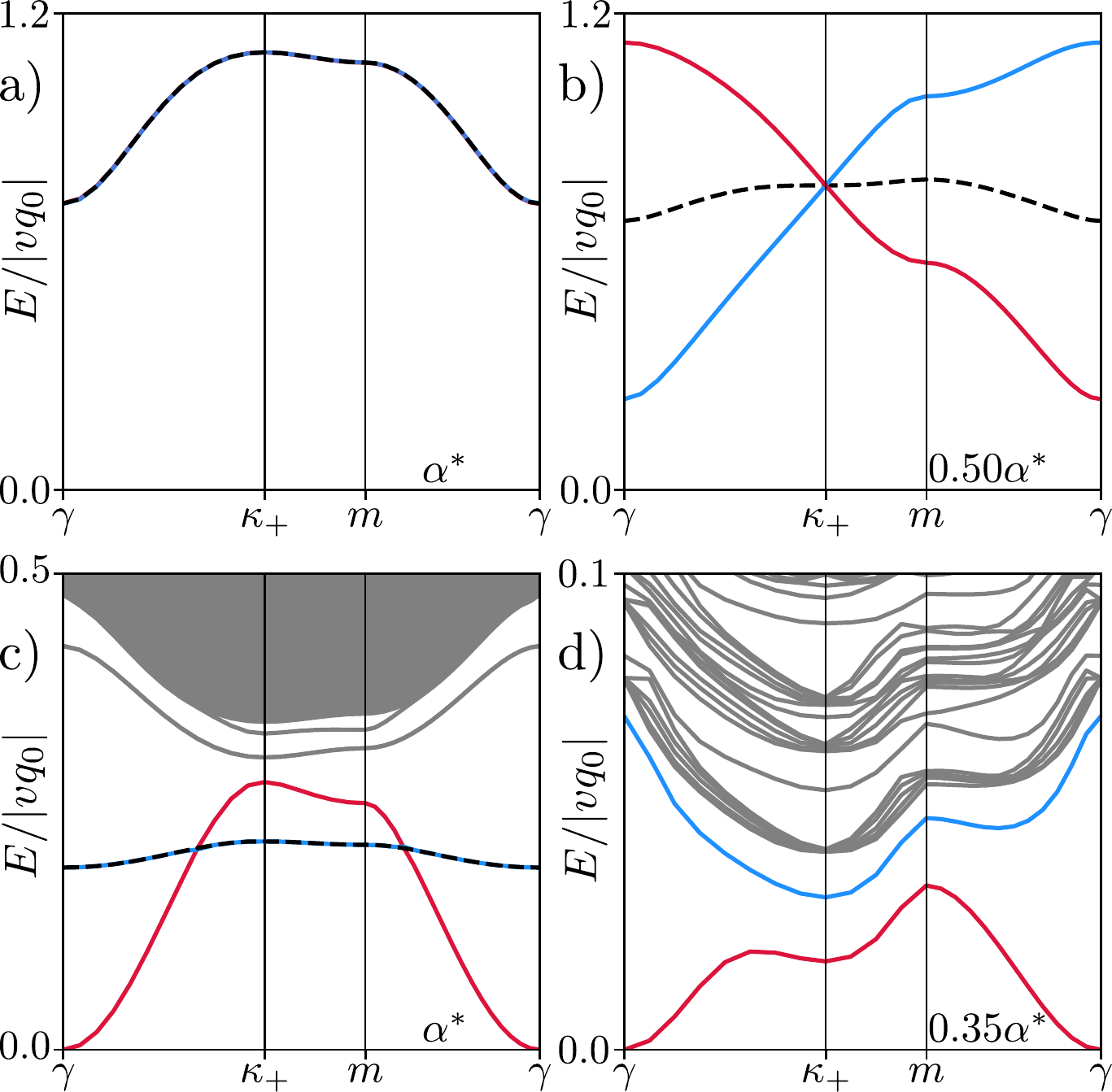}
\caption{(a-b) Charged and (c-d) neutral excitation spectrum above the state fully polarized along $\alpha = (\downarrow, -)$ at filling $n=-1$, at (a,c) and away from (b,d) the magic point. Charged excitation spectra (a,b) remain gapped over a wide region around $\alpha^*$, granting topological protection to the Chern insulator. The neutral excitation spectrum features a gapless magnon (red) and spin-degenerate gapped modes (blue, black) expected from the spontaneously broken SU(2) spin and $\mathbb{Z}_2$ chiral symmetries, respectively.
}
\label{fig:excitations}
\end{figure}

Figs.~\ref{fig:excitations}a-b show the spectra of charge-$e$ excitations above the ground state polarized along $\alpha = (\downarrow, -)$ at $n=-1$ (details in Appendix~\ref{app_chargegap}). The charge sector remains gapped over a rather large range of $\alpha$ around the magic point, which ensures quantization of the transverse conductance at low temperature and the dissipationless flow of edge current~\cite{niu1985quantized}.

The neutral excitation spectrum, shown in Figs.~\ref{fig:excitations}c-d, contains a gapless magnon branch resulting from the spontaneously broken continuous SU(2) spin symmetry; this mode disperses quadratically~\cite{nielsen1976count,watanabe2012unified}.
In addition, the broken discrete $\mathbb{Z}_2$ chiral symmetry yields a branch of gapped excitations with chirality opposite to that of the ground state.
We identify these states using a flavor-flip analysis (see Appendix~\ref{app_flavorflip}).

For $\alpha<\alpha^*$, the particle-hole dispersion softens the magnon spectrum near the $\kappa_\pm$ points, while the charge gap decreases. At a critical value of $\alpha$, which depends on the global interaction scale, excitations from both spectra fall below zero energy, which makes the ferromagnetic state unstable. The position of the collapse for the charge-$e$ spectrum is well captured by a Stoner-like criterion: it roughly occurs when the mean kinetic energy $\pm |W|$ of the bands near charge neutrality (Fig.~\ref{fig:BandStructure}) equals the charge gap obtained at $\alpha=\alpha^*$.
Using the parameters in Fig.~\ref{fig:excitations} as an illustration, the charge gap is equal to $\simeq 0.7 |vq_0|$ at the magic point, and it closes when $W = 0.66 |vq_0|$, which corresponds to $\alpha \simeq 0.3 \alpha^*$. Despite its smaller gap at $\kappa_+$ when $\alpha = \alpha^*$, the magnon spectrum closes at nearly the same value of $\alpha \simeq 0.32 \alpha^*$.

The phenomenology of these excitations reproduce that obtained in the chiral model for TBG~\cite{bernevig2021twistedI,song2021twisted,bernevig2021twistedIII,lian2021twisted,bernevig2021twisted,xie2021twisted}, which is expected from the exact mapping presented in Sec.~\ref{ssec:mappingchiralmodel}. Naturally, excitations above other insulating states, for instance the one stabilized at $n=0$ filling, can similarly be obtained. The only difference between TBG and the system studied here is the lower number of flavor-flip excitations available in the half-chiral model. As a result, some of the gapless excitations identified in TBG, such as valley-skyrmions, do not appear in our model.

\section{Material realization} \label{sec_symbreak}

We now discuss how to impose the periodic potential in Eq.~\eqref{eq_nearroot3matched} on graphene. Since the potential varies on a microscopic scale, it is unlikely to be realized by artificial patterning \cite{forsythe2018band,li2021anisotropic,barcons2022engineering}, 
which provides a valuable tuning knob for designer flat bands with a larger periodicity~\cite{shi2019gate,cano2021moire,wang2021moire,guerci2022designer,ghorashi2022topological,ghorashi2022multilayer}.
Instead, inspired by recent experiments on SiC/graphene heterostructures~\cite{lu2022dirac}, we propose to realize the model in a heterostructure where graphene sits on a substrate nearly lattice-matched to a $\sqrt{3} \times \sqrt{3}$ graphene supercell.
In Sec.~\ref{sec:candidates} we propose candidate materials to realize this situation.

In such a setup, the potential on graphene will generally contain additional terms beyond the leading harmonic of the chiral-symmetric model in Eq.~\eqref{eq_nearroot3matched}.
Sec.~\ref{ssec_othersubstrateterms} is devoted to considering such terms.
First, in Sec.~\ref{ssec_symrespecting}, we consider terms that break the chiral symmetry but preserve the crystalline symmetries of graphene. 
In presence of these terms, magic values of $V_1$ where the Fermi velocity vanishes still exist, but do not necessarily feature perfect flat bands. 
Then, in Sec.~\ref{ssec_symbreaking}, motivated by the material candidates in Table~\ref{tab:Materials}, we consider additional perturbations breaking some rotational and mirror symmetries of the half-chiral model.

\subsection{Candidate substrates}
\label{sec:candidates}

In Table~\ref{tab:Materials} and Appendix~\ref{Apx:mat_tables} we list candidate substrates whose surface periodicity nearly matches the $\sqrt{3} \times \sqrt{3}$ graphene supercell. As our proposed half-chiral model breaks translation symmetry in monolayer graphene via a potential term, but does not have interlayer hopping, we require candidate substrates to possess a sizeable band gap ($> 1$ eV). Table~\ref{tab:Materials} lists candidates predicted to be thermodynamically stable~\cite{doi:10.1021/acsami.6b01630,PhysRevB.85.235438} whose surface lattice vectors are near 2\% mismatch from the $\sqrt{3} \times \sqrt{3}$ graphene supercell.

The candidate substrates were identified from a pool of 2013 systems using the Materials Project API \cite{doi:10.1063/1.4812323,ONG2015209} to filter for those with reported experimental data, a band gap greater than 1 eV, and fewer than 10 sites per unit cell. Following the lattice matching procedure of Ref.~\cite{doi:10.1063/1.333084}, 90 candidate substrates were identified with hexagonal surface periodicity nearly matching the $\sqrt{3} \times \sqrt{3}$ graphene supercell. The 65 thermodynamically stable systems are listed in Table~\ref{tab:Materials} and Table~\ref{tab:other_stable} of Appendix~\ref{Apx:mat_tables}. Those in Table~\ref{tab:Materials} have a computed lattice mismatch of 1.5-2.5\%, yielding a \mr length scale similar to that of magic angle TBG. The remaining 25 candidates in Table~\ref{tab:metastable} are predicted to lie above the convex hull of the zero temperature Gibbs free energy and thus are metastable or only stable under particular conditions such as high pressure. These systems and their energies above the convex hull are tabulated in Table~\ref{tab:metastable} of
Appendix~\ref{Apx:mat_tables}.

The candidates we propose meet some necessary, but not sufficient, conditions for a physical realization of the emergent flat bands in the half-chiral model. 
They thus provide a good starting point for future theoretical and experimental investigation. 

One of many factors to determine whether graphene on a particular substrate realizes the half chiral model is the physical potential induced by that substrate. 
For each candidate, we have used the lattice mismatch $\epsilon$ to find $V_1^* = \alpha^* v \epsilon |K_0|$, shown in Table~\ref{tab:Materials}.
In practice, the full potential has a particular value of the first harmonic $V_1$, as well as higher harmonics. 
Their magnitudes depend on the particular termination of the substrate and on the relaxation of both graphene and the substrate surface. Tuning this potential via, for instance, substrate strain, may provide a viable path toward engineering flat bands.

\begin{table*}[t]
\centering
\begin{tabular}{|lll|lcc|ccc|c|}
\hline
 & Composition & mp-id & Surface & Lattice & SG & $|\epsilon|$\% & $V_1^*$ (meV) & \Mr length (nm) & Gap (eV) \\
\hline
\hline
 & CdS & \href{https://materialsproject.org/materials/mp-672?material_ids=mp-672}{672} & 001 & hex. & $P6_3mc$ & 1.56 & 102 & 15.84 & 1.12 \\
\cline{1-10}
 & CsLaS$_2$ & \href{https://materialsproject.org/materials/mp-561586?material_ids=mp-561586}{561586} & 111 & rhomb. & $R\bar{3}m$ & 1.59 & 104 & 15.54 & 2.59 \\
\cline{1-10}
 & TbTlSe$_2$ & \href{https://materialsproject.org/materials/mp-569507?material_ids=mp-569507}{569507} & 111 & rhomb. & $R\bar{3}m$ & 1.61 & 106 & 15.28 & 1.45 \\
\cline{1-10}
 & Ba$_2$CuBrO$_2$ & \href{https://materialsproject.org/materials/mp-552934?material_ids=mp-552934}{552934} & 111 & rhomb. & $R\bar{3}m$ & 1.62 & 106 & 15.23 & 2.33 \\
\cline{1-10}
 & BaHgO$_2$ & \href{https://materialsproject.org/materials/mp-3915?material_ids=mp-3915}{3915} & 111 & rhomb. & $R\bar{3}m$ & 1.65 & 108 & 14.95 & 2.37 \\
\cline{1-10}
 & NaGdSe$_2$ & \href{https://materialsproject.org/materials/mp-999489?material_ids=mp-999489}{999489} & 111 & rhomb. & $R\bar{3}m$ & 1.69 & 111 & 14.60 & 1.63 \\
\cline{1-10}
 & CsSmS$_2$ & \href{https://materialsproject.org/materials/mp-9082?material_ids=mp-9082}{9082} & 111 & rhomb. & $R\bar{3}m$ & 1.71 & 113 & 14.41 & 2.24 \\
\cline{1-10}
 & Ba$_2$BrN & \href{https://materialsproject.org/materials/mp-1018098?material_ids=mp-1018098}{1018098} & 111 & rhomb. & $R\bar{3}m$ & 1.80 & 118 & 13.74 & 1.31 \\
\cline{1-10}
 & YTlSe$_2$ & \href{https://materialsproject.org/materials/mp-1067744?material_ids=mp-1067744}{1067744} & 111 & rhomb. & $R\bar{3}m$ & 1.88 & 124 & 13.11 & 1.45 \\
\cline{1-10}
 & PrTlSe$_2$ & \href{https://materialsproject.org/materials/mp-999289?material_ids=mp-999289}{999289} & 111 & rhomb. & $R\bar{3}m$ & 1.89 & 124 & 13.06 & 1.45 \\
\cline{1-10}
 & RbBiS$_2$ & \href{https://materialsproject.org/materials/mp-30041?material_ids=mp-30041}{30041} & 111 & rhomb. & $R\bar{3}m$ & 1.92 & 126 & 12.89 & 1.37 \\
\cline{1-10}
 & DyTlSe$_2$ & \href{https://materialsproject.org/materials/mp-568062?material_ids=mp-568062}{568062} & 111 & rhomb. & $R\bar{3}m$ & 1.98 & 130 & 12.44 & 1.43 \\
\cline{1-10}
 & NdTlS$_2$ & \href{https://materialsproject.org/materials/mp-3664?material_ids=mp-3664}{3664} & 111 & rhomb. & $R\bar{3}m$ & 1.99 & 131 & 12.42 & 1.74 \\
\cline{1-10}
 & YbSe & \href{https://materialsproject.org/materials/mp-286?material_ids=mp-286}{286} & 111 & cubic & $Fm\bar{3}m$ & 2.08 & 136 & 11.89 & 2.03 \\
\cline{1-10}
 & SrHClO & \href{https://materialsproject.org/materials/mp-24066?material_ids=mp-24066}{24066} & 001 & hex. & $P6_3mc$ & 2.09 & 137 & 11.82 & 4.98 \\
\cline{1-10}
 & HoTlSe$_2$ & \href{https://materialsproject.org/materials/mp-569178?material_ids=mp-569178}{569178} & 111 & rhomb. & $R\bar{3}m$ & 2.16 & 142 & 11.43 & 1.42 \\
\cline{1-10}
 & ZnTe & \href{https://materialsproject.org/materials/mp-2176?material_ids=mp-2176}{2176} & 111 & cubic & $F\bar{4}3m$ & 2.28 & 150 & 10.83 & 1.08 \\
\cline{1-10}
 & NaTbSe$_2$ & \href{https://materialsproject.org/materials/mp-999127?material_ids=mp-999127}{999127} & 111 & rhomb. & $R\bar{3}m$ & 2.35 & 154 & 10.52 & 1.90 \\
\cline{1-10}
 & RbLuSe$_2$ & \href{https://materialsproject.org/materials/mp-10785?material_ids=mp-10785}{10785} & 111 & rhomb. & $R\bar{3}m$ & 2.42 & 159 & 10.21 & 2.11 \\
\cline{1-10}
 & TiCoSb & \href{https://materialsproject.org/materials/mp-5967?material_ids=mp-5967}{5967} & 111 & cubic & $F\bar{4}3m$ & 2.42 & 159 & 10.20 & 1.07 \\
\cline{1-10}
 & Ca(MgAs)$_2$ & \href{https://materialsproject.org/materials/mp-9564?material_ids=mp-9564}{9564} & 001 & hex. & $P\bar{3}m1$ & 2.46 & 162 & 10.02 & 1.26 \\
\cline{1-10}
 & MnI$_2$ & \href{https://materialsproject.org/materials/mp-28013?material_ids=mp-28013}{28013} & 001 & hex. & $P\bar{3}m1$ & 2.49 & 163 & 9.93 & 1.18 \\
\cline{1-10}
\end{tabular}
\caption{
Thermodynamically stable candidate substrates with $1.5\% < |\epsilon| < 2.5\%$. Additional candidates are given in Appendix~\ref{Apx:mat_tables}. The columns, from left to right, indicate the stoichiometry of the substrate, the database ID (with hyperlink) of the corresponding entry in the Materials Project database \cite{doi:10.1063/1.4812323}, Miller indices of the particular surface, the bulk lattice type and space group, the magnitude of the \mr reciprocal lattice vector as a percentage of that of graphene $|\epsilon|\%$, the $V_1^*$ value corresponding to the mismatch, the moir\'e length scale, and the bulk band gap of the substrate.
}
\label{tab:Materials}
\end{table*}

\subsection{Substrate-induced symmetry breaking} \label{ssec_othersubstrateterms}

The potential induced on graphene from proximity to a substrate with a $\sqrt{3} \times \sqrt{3}$ supercell will generally contain additional terms beyond those in Eq.~\eqref{eq_nearroot3matched}. As in Ref.~\cite{scheer2023kagome}, we expand them as a function of reciprocal lattice vectors of the substrate. Projected into the low-energy basis given in Eq.~\eqref{eq_tbgbasis} (TBG$_K$ column in Table~\ref{tab:Symmetries}), the most general Hamiltonian for the graphene layer compatible with the moir\'e periodicity is a sum of harmonics
\begin{equation}\label{eq_symm_allowed_ham}
H_\text{eff}=\begin{bmatrix}\sum_G H_{G,+}e^{iG\cdot r}&\sum_Q H_Qe^{iQ\cdot r}\\\sum_Q H_Q^\dagger e^{-iQ\cdot r}&\sum_G H_{G,-}e^{iG\cdot r}\end{bmatrix}
\end{equation}
where $G$ runs over the moir\'e reciprocal lattice vectors and $Q$ runs over $G+q_0$. Eq.~\eqref{eq_symm_allowed_ham} simplifies to the half-chiral model in Eq.~\eqref{eq_originalcontinuummodel} when all sums are restricted to the lowest symmetry-related set of harmonics with $H_{G=0,\pm}$ the $k\cdot p$ Hamiltonian of graphene close to the $\pm K$ valley, and $H_{q_{0,1,2}}$ reproducing $T(r)$ in Eq.~\eqref{eq_TBGbasisTunneling}.

We now consider the leading corrections beyond the half-chiral limit. We work to zeroth order in $k$, aside from $H_{G,\pm}$ for $G=0$, which is computed to first order in $k$. 
Specifically, we first assume that crystal symmetries are preserved, but break chiral symmetry. We then characterize the behavior of the magic coupling, defined as the coupling for which the Fermi velocity vanishes, and the bandwidth at that coupling. Finally, since many of the substrates in Table~\ref{tab:Materials} break the rotational symmetry of graphene, we consider corresponding symmetry-breaking terms.
The derivation of the symmetry-breaking terms from symmetry considerations is performed in \Cref{Apx:SymmHamTermsDerivation}.

\subsubsection{Breaking chiral symmetry}\label{ssec_symrespecting}

To linear order in $k$, the most general form of the zeroth harmonic, $H_{0,\pm}$ allowed by crystal symmetry is
\begin{equation} \label{eq_zerothharmonicswithsymmetries}
H_{0,\pm}=\mu+vk\cdot\tau,
\end{equation}
which differs from Eq.~\eqref{eq_tbgbasis} by a renormalized Fermi velocity and an overall chemical potential, setting the scale and zero of energy, respectively. The latter can be physically tuned by electrostatic gating. 

New terms arise from the off-diagonal coupling $H_Q$ in Eq.~\eqref{eq_symm_allowed_ham}. The first and third harmonics, with $Q_i=\epsilon K_i$ and $Q_i=-2\epsilon K_i$, respectively, are constrained by crystal symmetry to the form
\begin{equation}\label{eq_HQ}
H_Q=V_n(\hat Q\times\tau)\cdot\hat z+it_n\tau_z,
\end{equation}
where $n=1,3$ specifies the harmonic and $V_n$, $t_n$ are real. The model Hamiltonian in Eq.~\eqref{eq_tbgbasis} only contains $V_1$ with no $t_n$.

The most general crystal-symmetry-allowed form for the second harmonic is
\begin{equation}\label{eq_HG}
H_{G,\pm}=V_2\pm t_{\perp,2}(\hat G\times \tau)\pm it_{z,2}\tau_z,
\end{equation}
where $V_2$, $t_{\perp,2}$, and $t_{z,2}$ are real.
The $t_{z,2}$ term takes opposite sign for $G$ and $-G$, whereas all other contributions take the same sign for both. None of these terms appear in the original half-chiral model.

\begin{figure*}
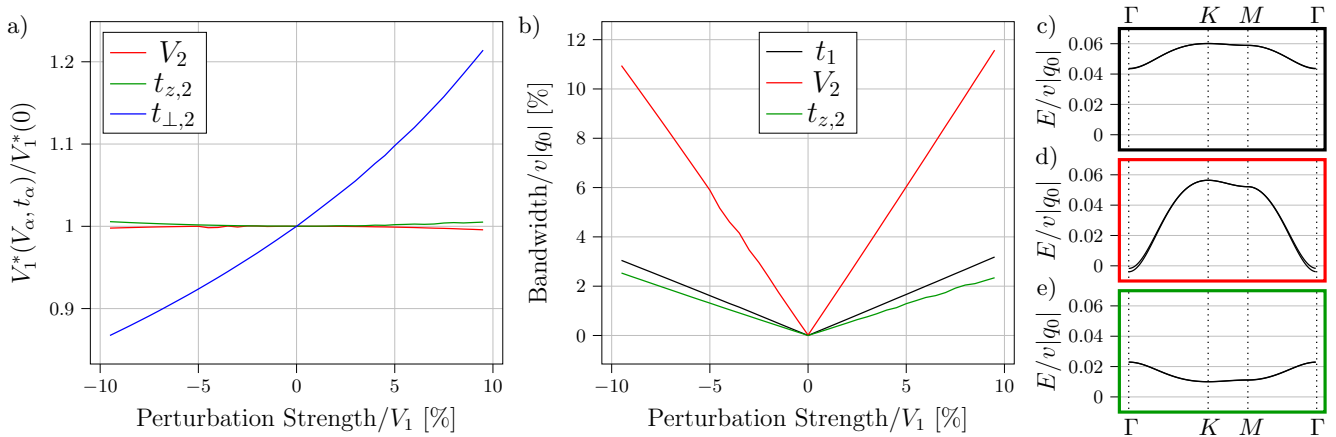

    \centering
    \begin{tikzpicture}[scale=0.8]
\input{V0MainHorizontal}
\node at (-1.2,5.6) {a)};
\begin{scope}[shift={(8.5,0)}]
\input{BandwidthMainHorizontal}
\node at (-1.2,5.6) {b)};
\begin{scope}[shift={(8.6,0)}]
\input{PerturbationSpectraStacked}
\end{scope}
\end{scope}
\end{tikzpicture}
    \caption{Corrections to the half-chiral model induced by small symmetry-preserving perturbations of the lowest two harmonics. (a) Only the $t_{\perp,2}$ term (blue) significantly changes the Fermi-velocity-minimizing value of $V_1^*$. The next-largest corrections come from $V_2$ (red) and $t_{z,2}$ (green). (b) Bandwidth as a function of $t_1$ (black), $t_{z,2}$ (green), and $V_2$ (red). (c-e) Representative spectra for those perturbations which induce bandwidth: (c) $t_1$, (d) $V_2$, and (e) $t_{z,2}$.
    }
    \label{fig:ChiralSymmBreakPertEffects}
\end{figure*}

In the presence of these additional terms, the Fermi velocity can always be tuned to zero by a single tuning parameter, which may physically correspond to lattice mismatch or pressure~\cite{FlatBandsFromSymmetries}. In general, the magic value of $V_1$ will be changed. Our numerical calculations of the band structure show that the only parameter which significantly changes the magic value is $t_{\perp,2}$ which adds a linear order correction to $V_1^*$, as shown in Fig.~\ref{fig:ChiralSymmBreakPertEffects}a. The other chiral-symmetry-breaking terms add quadratic order corrections. 

Moreover, terms which do not break the chiral symmetry (in these lowest few harmonics, $t_{\perp,2}$ and $V_3$) cannot break the perfect flatness of the bands, according to the symmetry arguments given in Ref.~\cite{FlatBandsFromSymmetries}. 
Terms that break chiral symmetry, however, generally introduce a band dispersion, as shown in Fig.~\ref{fig:ChiralSymmBreakPertEffects}b. Representative spectra are shown in Fig.~\ref{fig:ChiralSymmBreakPertEffects}c.

\subsubsection{Crystal symmetry breaking} \label{ssec_symbreaking}

In realistic implementations, many of the crystal symmetries of graphene will generically be broken by the substrate. In this section, motivated by our material candidates in \Cref{tab:Materials}, we will assume that SU(2)$_\text{spin}$, time reversal, and $C_{3z}$ symmetry are preserved.

\begin{figure}
    \centering
    \includegraphics[]{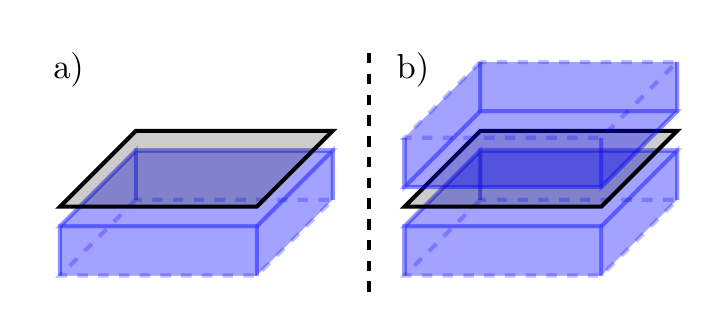}
    \caption{(a) Graphene on a substrate manifestly breaks the out of plane rotations $C_{2x}$ and $C_{2y}$. (b) A sandwiched heterostructure can preserve $C_{2x}$ or $C_{2y}$ symmetry.
    }
    \label{fig:StackSandwich}
\end{figure}

The remaining crystal symmetries are $M_z$, the twofold rotations $C_{2x,y,z}$, the in-plane mirror symmetries $M_{x,y}$, and spatial inversion $I$. The last six can be broken into two classes: those which preserve the $z$-orientation ($M_x$, $M_y$, and $C_{2z}$), and those which flip it ($C_{2y}$, $C_{2x}$, and $I$). Each term in the second class is the product of an element in the first class and $M_z$.
The latter class of symmetries is necessarily broken when graphene sits on top of the substrate, as shown in Fig.~\ref{fig:StackSandwich}a. If instead graphene is sandwiched between two symmetry-related substrate layers, as shown in Fig.~\ref{fig:StackSandwich}b, exactly one of these symmetries can be restored.

We now look to the physical symmetries of the substrates in \Cref{tab:Materials}. Analyzing the symmetry groups reveals that all of our material candidates have $M_x$, the reflection sending $x \mapsto -x$. In principle, it is always possible to restore both $M_x$ and $C_{2x}$ (and therefore $I=C_{2x}M_x$) in a sandwich structure. Therefore, in such a setup, none of the crystal symmetries in the effective model are broken.

However, since the sandwich structure may be difficult to achieve, and, further, makes it challenging to probe the embedded graphene layer, we restrict ourselves to studying the stacked structure shown in Fig.~\ref{fig:StackSandwich}a. Thus, we will consider $C_{2y}$, $C_{2x}$ and inversion symmetry to be broken. Since those substrates also lack $M_y$ and $C_{2z}$, we now consider all perturbations compatible with SU(2)$_\text{spin}$, $C_{3z}$, $T$, and $M_x$ symmetries.

For the first and third harmonics, this extension is remarkably simple: the coefficients $V_{1,3}$ and $t_{1,3}$ are allowed to become complex. For the second harmonic, a new term $H_{G,\pm}=\pm it_{||,2}(\hat G\cdot\tau)$ is allowed, where $t_{||,2}$ is real and identical for $G$ and $-G$.

Suppose one considers $C_{2z}$-breaking in the lowest-harmonic onsite potential, which manifests as a phase on $V_1$ (with other perturbations vanishing). This arises when the potential in Eq.~\eqref{eq_nearroot3matched} has a sine as well as a cosine component. The complex phase on $V_1$ can be eliminated by a basis change $\mathcal{U}_0=e^{i\theta\mu_z/2}$, where $\theta$ is the phase on $V_1$, which yields exactly the form in Eq.~\eqref{eq_originalcontinuummodel}. Thus, the mapping of the near-$\sqrt{3}$ matched potential model to chiral TBG is still valid with a sine component in the potential of Eq.~\eqref{eq_nearroot3matched}.

In fact, our numerical calculations show that \textit{none} of these $C_{2x}$-breaking perturbations induce a significant bandwidth or a significantly change in the magic value of $V_1$. 
This lack of bandwidth is not implied by the symmetry arguments of Ref.~\cite{FlatBandsFromSymmetries}.

\section{Conclusion} \label{sec_outlook}

In this manuscript, we have comprehensively studied a realization of the half-chiral model by imposing a near-$\sqrt{3}$ matched superlattice potential on a graphene monolayer. The half-chiral model reproduces most of the single-particle and interacting physics of the chiral model for TBG, albeit with half the degrees of freedom. The main difference between the half-chiral model presented here and the chiral model of TBG is in the physical properties of the excitations. For instance, valley-skyrmions are absent in our half-chiral model. Consequently, a realization of the half-chiral model may be a good platform to test which phases of TBG rely on the existence of such skyrmions. 

While preliminary data on a SiC/graphene heterostructure has already substantiated the relevance of our model~\cite{lu2022dirac}, reaching the flat band limit will require a stronger coupling, a closer lattice match, and mitigating symmetry breaking effects. In Table~\ref{tab:Materials} we propose several other material candidates. An important future question is whether these materials have potential to realize the flat band limit.

Our work will catalyze further design of novel heterostructures realizing analytically or numerically tractable moir\'e models.
Understanding these systems provides insight into the many-body physics of TBG.

\section{Acknowledgements}

We acknowledge insightful discussions with B. Lian, C. R\'epellin, and M. Scheer. V.C. is also grateful to N. Regnault for reviewing results from the TBG series~\cite{bernevig2021twistedI,song2021twisted,bernevig2021twistedIII,lian2021twisted,bernevig2021twisted,xie2021twisted}.
J.B., J.C., V.C., and D.G. acknowledge support from the Flatiron Institute, a division of the Simons Foundation.
Work by A.D. is supported by the National Science Foundation under the Columbia MRSEC on Precision-Assembled Quantum Materials (PAQM), Grant No. DMR-2011738.
J.C. acknowledges the Alfred P. Sloan Foundation through a Sloan Research Fellowship and hospitality by the Kavli Institute for Theoretical Physics, where this research was supported in part by the National Science Foundation under Grant No. NSF PHY-1748958.

\bibliography{BibChiral}

\appendix 

\section{Projection of onsite potential}\label{Apx:basics}

In this appendix, we derive the projected Hamiltonian in Eq.~\eqref{eq_originalcontinuummodel} from Eqs.~\eqref{eq_bareHKKp} and \eqref{eq_nearroot3matched}.
For convenience, we rewrite the potential in Eq.~\eqref{eq_nearroot3matched} here as
\begin{equation}
    V( r)=V_1\sum_n \left[e^{iK_n\cdot r}e^{i\epsilon K_n\cdot r}+c.c.\right].
\end{equation}
We define the Bloch basis:
\begin{equation}
    \psi_{k,A/B}(r)=\frac{1}{\sqrt{N}}\sum_R\varphi(r-R_{A/B})e^{ik\cdot R_{A/B}}
\end{equation}
where $R_{A/B}=R\pm u$, $R$ is the lattice vector of graphene, $N$ the number of unit cells, $u=(r_A-r_B)/2=a(1,0)$ and $\varphi(r-R_{A/B})$ is the $p_z$ orbital of the carbon atoms. Projecting the potential on the Bloch basis we find 
\begin{equation}
\label{app:matrix_el_1}
    V_{\alpha,\beta}(k,k')=\frac{\delta_{\alpha\beta}}{N}\sum_R V(R +u_\alpha) e^{-i(k-k')(R+u_\alpha)},
\end{equation}
where $\alpha=A,B$, $u_{A/B}=\pm u$ and we have employed the relation 
\begin{equation}
    \int d^2 {\bf r} \varphi^*(r-R_\alpha)V(r)\varphi(r-R_\beta)=\delta_{\alpha\beta} V(R_\alpha).
\end{equation}
We first observe that the wave vectors $K_n$ in $V(r)$ connects the two Dirac cones of graphene, $K-K'=K_n+G_n$ with $G=[0,-e_-,-e_+]$ for $n=0,1,2$. The long wavelength modulation $q_n=\epsilon K_n$ gives rise to the \mr lattice in Fig.~\ref{fig:ModelAndMoire} and introduces the momentum selection rule $\bar k-\bar k'=q_n$ where $\bar k$ and $\bar k'$ are small variations around $K$ and $K'$. As a result, in the sublattice basis $[A,B]$ the matrix element~\ref{app:matrix_el_1} takes the form
\begin{equation}
    V_{K,K'}(\bar k,\bar k')=V_1\sum^2_{n=0}\delta_{\bar k-\bar k',q_n}
    \begin{bmatrix}
    e^{- i u\cdot G_n} & 0\\ 0 & e^{ i u\cdot G_n}
    \end{bmatrix},
\end{equation}
where we defined $V_{K\alpha,K'\beta}(\bar k,\bar k')\equiv V_{\alpha,\beta}(K+\bar k,K'+\bar k')$. Noticing that $e_+=K_0-K_1$ and $g_-=K_0-K_2$ we find $e_+\cdot u=-K_1\cdot u=-\phi=2\phi$ and $e_-\cdot u=-K_2\cdot u=\phi$ where we used the relation $K_n\cdot(r_A-r_B)=2K_n\cdot u=2n\phi$. Finally, we find
\begin{equation}
V_{ K, K'}( \bar k, \bar k')=V_1\sum_n \delta_{ \bar k, \bar k'+q_n} \begin{bmatrix}
e^{in\phi} & 0\\ 0 & e^{-in\phi}
\end{bmatrix},    
\end{equation}
which in real space becomes Eq.~\eqref{eq_originalcontinuummodel}.

\section{Charge gap} \label{app_chargegap}

In this appendix, we study the charged excitations above the ferromagnetic ground state at filling $n=-1$.
We choose the ground state to be polarized in the $\alpha = (-,\downarrow)$ flavor, $\ket{{\rm FM}} = \prod_{k} c_{k,-,\downarrow}^\dagger \ket{\emptyset}$, with $\ket{\emptyset}$ the state with all bands empty (filling $n=-2$). The charged excitations are accompanied by either a spin flip, a chirality flip, or both:
\begin{equation} \label{eq_singlecharged} \begin{split}
\ket{\phi_S(k)} &= c_{k,-,\uparrow}^\dagger \ket{{\rm FM}} , \\
\ket{\phi_\Lambda(k)} &= c_{k,+,\downarrow}^\dagger \ket{{\rm FM}} , \\
\ket{\phi_T(k)} &= c_{k,+,\uparrow}^\dagger \ket{{\rm FM}} .
\end{split}
\end{equation}
Since the Hamiltonian is translation invariant, the momentum $k$ of these states is conserved and finding the dispersion of these charged excitations only involves diagonalizing a 3$\times$3 matrix.

Away from $\alpha=\alpha^*$, the band and chirality indices are no longer identical and the band eigenmodes $d_{k, n = \pm, \sigma}$ with energy $n \varepsilon_k$, can be expressed as linear combinations of the chiral fermionic operators $c_{k, \gamma=\pm, \sigma}$. Since $\Lambda_z$ anticommutes with the Hamiltonian, the dispersion must be off-diagonal when written in the chiral basis
\begin{equation}
\label{H_chiral_kinetic}
H_{k} = \sum_{k,n,\sigma} n \varepsilon_{k} d_{n,\sigma}^\dagger d_{n,\sigma} = \sum_{k,\sigma} [ u_{k} c_{k,+,\sigma}^\dagger c_{k,-,\sigma} + hc] ,
\end{equation}
where the $u_{k}$ are determined numerically by projection. Finally, recall that the interaction
\begin{equation}
H_{\rm int} = \frac{1}{2} U_{(\alpha,\beta)} (k_1, k_2,q) c_{k_1,\alpha}^\dagger c_{k_1+q,\alpha} c_{k_2+q,\beta}^\dagger  c_{k_2,\beta}  ,
\end{equation} 
has diagonal form factors in the chiral basis 
\begin{equation}
U_{(\alpha,\beta)} (k_1, k_2,q) = \sum_G U_{q+G} F_{k_1,q+G}^\alpha  F_{k_2,q+G}^{\beta \, *}  .
\end{equation}

Finding the dispersion of a charged excitation above the ferromagnet $\ket{{\rm FM}}$ amounts to diagonalizing the kinetic and interacting Hamiltonian in the $[\phi_\Lambda(k), \phi_T(k), \phi_S(k)]$ subspace, where the matrix elements read
\begin{equation}
H_k = \begin{bmatrix} I_{\overline{\rm FM}} (k) & 0 & 0 \\ 0 &  I_{\overline{\rm FM}} (k) & u_k \\ 0 & u_k^* &  I_{{\rm FM}} (k) \end{bmatrix} 
\end{equation}
and Wick contractions provide the simple form  
\begin{equation}
I_{M}(k) = \sum_q U_{M,{\rm FM}} (k,q,0) + U_{M,M}(k,k,q)/2 .
\end{equation}

Finally, since our flavor-flip analysis serves the purpose of illustrating the effects of interactions in our model, we model the electronic repulsion with a positive term $U_{q} = U >0$ for all $q$, \textit{i.e.} a local repulsion between all flavors in real-space. In Fig.~\ref{fig:excitations}, we have used an interaction amplitude $U = 0.5 |v q_0|$.

\section{Charge neutral excitations} \label{app_flavorflip}

\subsection{Setup}

The neutral excitations above the Chern insulating ground state at $n=\pm 1$ are of three different types: spin-flipping ($S$), chirality-flipping ($\Lambda$) and spin-chirality-flipping ($T$). They respectively take the form 
\begin{equation} \label{eq_singleflavorflip} \begin{split}
\ket{k, S}_Q &= c_{k+Q,-,\uparrow}^\dagger c_{k,-,\downarrow} \ket{{\rm FM}} , \\
\ket{k, \Lambda}_Q &= c_{k+Q,+,\downarrow}^\dagger c_{k,-,\downarrow} \ket{{\rm FM}} , \\
\ket{k, T}_Q &= c_{k+Q,+,\uparrow}^\dagger c_{k,-,\downarrow} \ket{{\rm FM}} .
\end{split}
\end{equation}
Since the Hamiltonian is translation invariant, the center of mass momentum $Q$ is conserved, and we shall leave it implicit when no confusion is possible. 

The well-defined spin $S_z$ and $\Lambda_z$ quantum number of the excitations in Eq.~\eqref{eq_singleflavorflip} only allow for a few non-zero elements of the kinetic Hamiltonian defined in Eq.~\eqref{H_chiral_kinetic}, which are given by
\begin{equation}
\braOket{k',S}{H_{\rm k}}{k,T} = \braOket{k,T}{H_{\rm k}}{k',S}^* =  \delta_{k',k} u_{k+Q} . 
\end{equation}

The matrix elements of $H_{\rm I}$ in the basis of single particle excitations Eq.~\eqref{eq_singleflavorflip} can be obtained using Wick contractions. In the chiral limit, where the form factor is diagonal in the chiral basis, as shown in Sec.~\ref{ssec_symfornfactors}, they only couple states with the same flavor-flip $\braOket{k',M}{ H_{\rm I} }{k,N} = \delta_{M,N} \braOket{k'}{ H_{\rm I} }{k}_M$, and take the following form
\begin{align}  \label{eq_interactionmatrixelement}
& \braOket{k'}{ H_{\rm I} }{k}_M = -  U_{M,{\rm FM}} (k+Q, k, k'-k)  \\ 
& \, +  \delta_{k',k} \textstyle{\sum_{\tilde k}} [U_{M, {\rm FM}} (k+Q, \tilde k, 0) -  U_{{\rm FM}, {\rm FM}} (k, \tilde k, 0)] \notag \\
& \, + \delta_{k',k} \textstyle{\sum_{\tilde k}} [ U_{{\rm FM}, {\rm FM}} (k, k, \tilde k) + U_{M,M} (k+Q, k+Q, \tilde k)  ] /2, \notag
\end{align}
which was decomposed into a direct electron-hole interaction (first line), an electron and hole Hartree contribution (second line), and a sum of Fock terms (third line). In this calculation, we have subtracted the interaction energy of the $\ket{\rm FM}$ state, as in Appendix~\ref{app_chargegap}, and used the relation $U_{\beta\alpha}(k_2+q,k_1+q,-q) = U_{\alpha\beta}(k_1,k_2,q)$ which only holds true for inversion symmetric Coulomb potentials $U_{-q} = U_q$.

Once these coefficients have been computed, collective excitations above $\ket{{\rm FM}}$ are obtained by diagonalizing
\begin{equation} \label{eq_flavorflipHamiltonian}
\braOket{k'}{ H_{\rm I} }{k}_{\overline{\rm FM}} , \quad \begin{bmatrix} \braOket{k'}{ H_{\rm I} }{k}_{\overline{\rm FM}} & {\rm diag} (u_{k+Q}) \\ {\rm diag} (u_{k+Q}^*) & \braOket{k'}{ H_{\rm I} }{k}_{\rm FM} \end{bmatrix} ,
\end{equation}
corresponding to the $\Lambda$ and $(S,T)$ subspaces. Note that each block represented by its $(k',k)$ elements is a matrix whose dimension equals the number of Brillouin zone points.

\subsection{Chiral limit}

To start with we consider the magic angle case $\alpha=\alpha^*$, for which $u_k = 0$, and where Fig.~\ref{fig:excitations}c shows a gapless quadratic branch and a few gapped modes below the many-body continuum. 
We can capture these features using a small $Q$ Taylor expansion. 

Let us first show that the lowest energy mode of the spin-flipping $S$ branch has exactly zero energy at $Q=0$ by diagonalizing the interaction term in the $M={\rm FM}$ sector. Clearly, the Hartree contribution at $Q=0$ vanishes in this sector (see Eq.~\eqref{eq_interactionmatrixelement}). While the diagonal Fock term is positive, it is precisely opposite to the sum of the direct interaction matrix elements along each row $\sum_{k'} \braOket{k'}{ H_{\rm I} }{k}_{\rm FM} = 0$. 
By the Gershgorin circle theorem, this ensures that all eigenvalues of the interaction in the ${\rm FM}$ sector are non-negative.  
The lowest possible value, zero, is reached by a vector $\vec{V}_{\rm mag}$ whose coefficients are all equal corresponding to the Goldstone mode associated with the breaking of the SU(2)$_\text{Spin}$ symmetry. 
In the $\overline{{\rm FM}}$ sector at $Q=0$ we have the additional positive Hartree term $\delta_{k',k} \textstyle{\sum_{\tilde k}} [U_{\overline{\rm FM}, {\rm FM}} (k, \tilde k, 0) -  U_{{\rm FM}, {\rm FM}} (k, \tilde k, 0)]$, see second line of Eq.~\eqref{eq_interactionmatrixelement}.
The latter introduces a gap $\Delta_n>0$ for the first few chiral-breaking excitations $\vec{V}_{\rm Z2}^{(n)}$ with $n = 1,2,3\cdots $ (in the $T$ and $\Lambda$ sectors); the subscript ${\rm Z2}$ refers to the chiral-flip character of these collective modes.

The Goldstone mode dispersion around $\gamma$ is quadratic in $Q$ as expected from general arguments~\cite{watanabe2012unified}. To prove the vanishing of the linear term we perform a $k \cdot p$ expansion around $\gamma$,
\begin{equation} \begin{split}
\epsilon_Q^{\rm mag} &\simeq Q \cdot [\vec{V}_{\rm mag}^\dagger (\nabla_Q H_{\rm I}) \vec{V}_{\rm mag}] \\ &= Q \cdot \nabla_Q \left[ \frac{1}{N_{BZ}}\sum_{k,k'} \braOket{k'}{H_{\rm I}}{k}_{\rm FM} \right] ,
\end{split} \end{equation}
with $N_{BZ}$ the number of point in the Brillouin zone. We now look at coefficient of the linear in momentum term:
\begin{widetext}
\begin{equation}
    \begin{split} 
    \label{linear_vanishing_1}
    &\frac{\nabla_Q}{N_{BZ}}\sum_{k,k'} \braOket{k'}{H_{\rm I}}{k}_{\rm FM}=\sum_{k,q}\sum_G \frac{U_{q+G}}{N^2_{BZ}}\Re\left[F^{-*}_{k,q+G}\left(\nabla_k F^-_{k,q+G}\right)\right]+\sum_G \frac{U_G}{N^2_{BZ}} \left[\sum_{p}\nabla_p F^-_{p,G}\right]\left[\sum_{k}F^{-*}_{k,G}\right]\\
    &-\sum_{k,k'}\sum_{G}\frac{U_{k-k'+G}}{N^2_{BZ}}F^{-*}_{k',k-k'+G}\left[\braket{\psi_{k',-}}{\nabla_k \psi_{k+G,-}}+\braket{\nabla_{k'}\psi_{k',-}}{ \psi_{k+G,-}}\right].
    \end{split}
\end{equation}
Making the change of variables $q=k'-k$ yields
\begin{equation}
    \begin{split} 
    \label{linear_vanishing_2}
    &\frac{\nabla_Q}{N_{BZ}}\sum_{k,k'} \braOket{k'}{H_{\rm I}}{k}_{\rm FM}=\sum_{k,q}\sum_G \frac{U_{q+G}}{N^2_{BZ}}\Re\left[F^{-*}_{k,q+G}\left(\nabla_k F^-_{k,q+G}\right)\right]-\sum_{k,q}\sum_{G}\frac{U_{q+G}}{N^2_{BZ}}F^{-*}_{k,q+G}\left(\nabla_k F^{-}_{k,q+G}\right)\\
    &+\sum_G \frac{U_G}{N^2_{BZ}} \left[\sum_{p}\nabla_p F^-_{p,G}\right]\left[\sum_{k}F^{-*}_{k,G}\right].
    \end{split}
\end{equation}
\end{widetext}
By sending $q\to-q$, $G\to-G$, using the relation $U_{q}=U_{-q}$ we readily find that the second contribution on the right hand side of Eq.~\eqref{linear_vanishing_2} is real and cancels the first one.   
We are left with 
\begin{equation}
    \frac{\nabla_Q}{N_{BZ}}\sum_{k,k'} \braOket{k'}{H_{\rm I}}{k}_{\rm FM}=\sum_{p,k,G} \frac{U_G}{N^2_{BZ}}(\nabla_p F^-_{p,G})F^{-*}_{k,G}.
\end{equation} 
To show the vanishing of the right hand side we first make the change of variable $p'=C_{3z}p $ so that $(\nabla_p F^-_{p,G})=(C_{3z}\nabla_{p'} F^-_{C^{-1}_{3z}p,G})$. Then, we use the $C_{3z}$ symmetry $F^{-}_{C_{3z}p,G}=F^{-}_{p,C^{-1}_{3z}G}$ and $U_{C_{3z}G}=   U_G$. Finally, we find:  
\begin{equation}
    \frac{\nabla_Q}{N_{BZ}}\sum_{k,k'} \braOket{k'}{H_{\rm I}}{k}_{\rm FM}=C_{3z}\sum_{p,k,G} \frac{U_G}{N^2_{BZ}}(\nabla_p F^-_{p,G})F^{-*}_{k,G},
\end{equation} 
i.e. $C_{3z} v=v$ with $v$ two-dimensional vector, which implies
\begin{equation}
    \frac{\nabla_K}{N_{BZ}}\sum_{k,k'} \braOket{k'}{H_{\rm I}}{k}_{\rm FM}=0.
\end{equation}
The dispersion of the Goldstone modes goes like $Q^2$ for small $Q$, i.e.,
\begin{equation}
\epsilon_Q^{\rm mag} = \frac{1}{2}m_{ij} Q_iQ_j 
\end{equation}
where
\begin{equation} \begin{split}
&m_{ij}=\vec{V}_{\rm mag}^\dagger (\partial_i\partial_j H_{\rm I}) \vec{V}_{\rm mag} \\  &-2\sum_{n\neq {\rm mag}} \frac{(\vec{V}^\dagger_{\rm mag}\partial_i H_{\rm I}\vec{V}^{(n)}_{Z2}) (\vec{V}^{(n)}_{Z2} \partial_j H_{\rm I}\vec{V}_{\rm mag}) }{\Delta_n}  ,
\end{split} \end{equation}
with $\partial_i=\partial_{Q_i}$ derivative with respect to the center of mass momentum $Q$ and $\Delta_n$ is the energy of the chiral-flip mode $\vec{V}^{(n)}_{Z2}$.

\subsection{Reference kinetic energy}

In this section we highlight the effect of finite dispersion away from the magic angle condition in the chiral limit. 
We now derive the kinetic energy gain of the ferromagnetic ground state. 
To find it, we note that the $\ket{{\rm FM}}$ ground state is connected to states in the $\Lambda$ sector at $Q=0$ through
\begin{equation}
\begin{bmatrix} \braOket{k'}{ H_{\rm I} }{k}_{\overline{\rm FM}} & u_k \\ u_{k'}^T & 0   \end{bmatrix} .
\end{equation}
It is important to notice that the latter Hamiltonian coincides precisely the matrix that couples $\vec{V}_{\rm mag}$ to the gapped excitation in the $\overline{\rm FM}$ sector.  
To second order in perturbation theory, we thus have
\begin{equation} \label{appeq_reductionFMenergy}
E_{\rm FM}^{\rm kin} \simeq - \sum_{n} \frac{|\braOket{\vec{V}_{\rm Z2}^{(n)}}{u_k}{\vec{V}_{\rm mag}}|}{\Delta_n} , 
\end{equation}
with $\Delta_n$ the energy of $\vec{V}_{\rm Z2}^{(n)}$, both for the ferromagnetic state and the $Q=0$ magnon state. In other words the hybridization induced by the finite bandwidth with the spin-chirality-flipping ($T$) modes gives rise to the same energy gain in the ferromagnetic ground state and in the $Q=0$ magnon state. 
This coincidence of energy level extends to all order in perturbation theory due to the SU(2) symmetry of the original model~\cite{lian2021twisted}.

Another way to understand the degeneracy is to momentarily forget to subtract off the ferromagnetic ground state energy. In that case, the energy reduction in Eq.~\eqref{appeq_reductionFMenergy} pulls the $Q=0$ magnon state below zero energy, hinting that it is the new ferromagnetic ground state of the system. However, $\vec{V}_{\rm mag}$ can be regarded as a spin-flip acting on all particles at the same time, or a global spin rotation of $\ket{\rm FM}$ itself. This would mean that a spin-rotation of the ferromagnet is more stable than the ferromagnet itself, in contradiction with the global SU(2) symmetry of the problem and its spontaneous symmetry breaking.

\section{Symmetry Analysis\label{Apx:SymmHamTermsDerivation}}
\begin{table}[]
\colorlet{shaded}{gray!60}
\centering
\begin{tabular}{|c|c|c||c|c||c|c|c|c|}
\hline
\multicolumn{5}{|c||}{Hamiltonian Form}&\multicolumn{4}{c|}{Symmetries}\\
\hline
Block & Coeff. & Term & SV & $-G$ & $\Lambda_z$ & $C_{2x}$ & $C_{2y}$ & $C_{2z}$ \\ \hline

\multirow{4}{3em}{$H_{\pm}$}
&$A$&$\tau_0$ &$+$&\cellcolor{shaded}& X & \checkmark & \checkmark & \checkmark\\
&$B$&$\tau_z$ &$-$&\cellcolor{shaded}& X & \checkmark & X & X\\ 
&$C$&$k\cdot\tau$ &$+$&\cellcolor{shaded}& \checkmark & \checkmark & \checkmark & \checkmark\\
&$D$&$k\times\tau$ &$+$&\cellcolor{shaded}& \checkmark & X & X & \checkmark\\ \hline

\multirow{6}{3em}{$H_Q$}
&$\Re{B}$&$\tau_z$ &\cellcolor{shaded}&\cellcolor{shaded}& X & $-_x$ & $+_x$ & X\\
&$\Im{B}$&$i\tau_z$ &\cellcolor{shaded}&\cellcolor{shaded}& X & $+_x$ & $+_x$ & \checkmark\\
&$\Re{C}$&$\hat Q\cdot\tau$ &\cellcolor{shaded}&\cellcolor{shaded}& \checkmark & $-_x$ & $-_x$ & \checkmark\\
&$\Im{C}$&$i\hat Q\cdot\tau$ &\cellcolor{shaded}&\cellcolor{shaded}& \checkmark & $+_x$ & $-_x$ & X\\
&$\Re{D}$&$(\hat Q\times\tau)\cdot\hat z$ &\cellcolor{shaded}&\cellcolor{shaded}& \checkmark & $+_x$ & $+_x$ & \checkmark\\
&$\Im{D}$&$i(\hat Q\times\tau)\cdot\hat z$ &\cellcolor{shaded}&\cellcolor{shaded}& \checkmark & $-_x$ & $+_x$ & X\\ \hline

\multirow{8}{3em}{$H_{G,\pm}$}
&$\Re{A}$&$\tau_0$ &$+$&$+$& X & $+_x$,$+_y$ & $+_x$,$+_y$ & \checkmark\\
&$\Im{A}$&$i\tau_0$ &$+$&$-$& X & $-_x$,$+_y$ & $+_x$,$-_y$ & X\\
&$\Re{B}$&$\tau_z$ &$-$&$+$& X & $+_x$,$+_y$ & $-_x$,$-_y$ & X\\
&$\Im{B}$&$i\tau_z$ &$-$&$-$& X & $-_x$,$+_y$ & $-_x$,$+_y$ & \checkmark\\
&$\Re{C}$&$\hat G\cdot\tau$ &$-$&$-$& \checkmark & $+_x$,$-_y$ & $+_x$,$-_y$ & \checkmark\\
&$\Im{C}$&$i\hat G\cdot\tau$ &$-$&$+$& \checkmark & $-_x$,$-_y$ & $+_x$,$+_y$ & X\\
&$\Re{D}$&$(\hat G\times\tau)\cdot\hat z$ &$-$&$-$& \checkmark & $-_x$,$+_y$ & $-_x$,$+_y$ & \checkmark\\
&$\Im{D}$&$i(\hat G\times\tau)\cdot\hat z$ &$-$&$+$& \checkmark & $+_x$,$+_y$ & $-_x$,$-_y$ & X \\ \hline
\end{tabular}
\caption[DummyText]{Terms that can be added with arbitrary real coefficients to the model Hamiltonian in Eq.~\eqref{eq_tbgbasis} in the TBG$_K$ basis while respecting time reversal, $C_{3z}$, and $SU(2)_\text{Spin}$ symmetries. 
The first five columns describe the additional term, as follows: The first column lists to which block of Eq.~\eqref{eq_symm_allowed_ham_apx} the term is added. The second column lists its coefficient in the notation of Eq.~\eqref{eq_C3z_constraint}, and the third column lists the term itself, for a $C_{3z}$-related set of $Q$s/$G$s.
The fourth column describes the relative sign between the two sublattice-valley combinations (the $\pm$ subscript in $H_{\pm}$ and $H_{G,\pm}$), determined by time-reversal symmetry. The fifth column describes the relative sign between $G$ and $-G$ imposed by hermiticity.
The last four columns indicate whether the term respects (\checkmark) or breaks (X) important symmetries of the model. 
For the $C_{2x}$ and $C_{2y}$ columns, $\pm_{x,y}$ indicates the relative sign of the coefficient when $Q$, $G$ undergo the reflections $M_{x,y}$.
}
\label{tab:GenHamTerms}
\end{table}

In this appendix, we derive the most general form of the low-energy Hamiltonian that respects time-reversal, $C_{3z}$, and $SU(2)_\text{Spin}$ symmetry. The resulting symmetry-allowed terms are summarized in \Cref{tab:GenHamTerms}.

We work in the TBG$_K$ basis because the form of $C_{3z}$ symmetry, $\exp(i\pi\tau_z/3)$, is valley-independent, which leads to more elegant expressions. We will also neglect spin entirely due to the assumed $SU(2)$ symmetry.

By arguments analogous to those given in \Cref{Apx:basics}, the low-energy effective Hamiltonian takes the form
\begin{equation}\label{eq_symm_allowed_ham_apx}
H_\text{eff}^{\text{TBG}_K}=\begin{bmatrix}\sum_G H_{G,+}e^{iG\cdot r}&\sum_Q H_Qe^{iQ\cdot r}\\\sum_Q H_Q^\dagger e^{-iQ\cdot r}&\sum_G H_{G,-}e^{iG\cdot r}\end{bmatrix}.
\end{equation}
$G$ runs over the moir\'e reciprocal lattice vectors $\epsilon g$ (where $g$ is a reciprocal lattice vector of graphene) and $Q$ runs over $G+\epsilon K_0$. In this way, at $\epsilon=0$, the sum over $G$ describes all harmonics which couple $K$ to itself, and the sum over $Q$ describes all harmonics that couple $K$ to $K'$.

We compute $H_{G=0,\pm}$ (abbreviated as $H_{\pm}$) to first order in $k$, and all other terms to zeroth order in $k$. The symmetry $C_{3z}=\exp(i\pi\tau_z/3)$ constrains each $2\times 2$ block of the Hamiltonian. The term $H_{G=0,\pm}$ can be written as
\begin{equation}\label{eq_C3z_constraint}
    H_\pm(k)=A\tau_0+B\tau_z+C(k\cdot \tau)+D(k\times\tau)\cdot\hat z.
\end{equation}
Other harmonics $H_{G,\pm}$ can be expanded with the same coefficients by making the replacement $k\rightarrow \hat G$, and similarly $H_Q$ with $k\rightarrow \hat Q$.

In this decomposition, the coefficients $A$-$D$ are the same for $C_{3z}$-related $Q$s and $G$s. The coefficients are generally complex, but hermiticity imposes the constraint $H_{G,\pm}=H_{-G,\pm}^\dagger$. Therefore, we can derive $A(G)=A(-G)^*$ and $B(G)=B(-G)^*$. Similarly, for $G\neq 0$, we find $C(G)=-C(-G)^*$ and $D(G)=-D(-G)^*$. For the special case $G=0$, hermiticity implies the coefficients $A$-$D$ of $H_{\pm}$ are real.

We now derive the constraints from time-reversal symmetry, chiral symmetry, and the remaining crystal symmetries on each of these terms individually; the form of each symmetry operator is listed in \Cref{tab:Symmetries}. Among these, the terms which respect time reversal are listed in \Cref{tab:GenHamTerms}, along with which of the other symmetries they respect.

Note that when considering the action of $M_x$ and $M_y$ constrained to vectors in the plane (such as $k$, $Q$, and $G$), we have $M_x=-M_y$, since $M_xM_y=C_{2z}=-I$.

\subsection{Zeroth harmonic: $H_{\pm}$}

We now consider how symmetry constrains the action of the zeroth harmonic $H_{\pm}$ in terms of the (real) coefficients $A$-$D$. Note the $\pm$ sign refers to the sublattice-valley (SV) index, as we are working in the TBG basis; accordingly, we will call the coefficients SV-even (resp. SV-odd) if they take opposite sign (resp. same sign) in the two valleys.

Time reversal relates the two valleys,
\begin{equation}
    H_\pm(k)=\tau_yH_\mp^*(-k)\tau_y,
\end{equation}
implying that $B$ must be SV-odd but the other coefficients must be SV-even. The anticommuting chiral symmetry requires
\begin{equation}
    H_\pm(k)=-\tau_zH_\pm(k)\tau_z,
\end{equation}
implying $A=B=0$. $C_{2x}T$ requires
\begin{equation}
    H_\pm(k)=\tau_zH_\pm(M_xk)^*\tau_z,
\end{equation}
implying $D=0$. $C_{2y}$ requires
\begin{equation}
    H_\pm(k)=\tau_yH_\pm(M_xk)\tau_y,
\end{equation}
implying $B=D=0$. Finally, $C_{2z}T$ requires
\begin{equation}
    H_\pm(k)=\tau_xH_\pm(k)^*\tau_x,
\end{equation}
implying $B=0$.

Altogether, the crystal symmetries restrict the form of the zeroth harmonic to $H_{\pm}=A+C (k\cdot\tau)$, which is used in Eq.~\eqref{eq_zerothharmonicswithsymmetries} of the main text.

\subsection{$K$-to-$K'$ harmonics: $H_Q$}

For the off-diagonal terms in Eq.~\eqref{eq_symm_allowed_ham_apx}, $H_Q$, time-reversal implies 
\begin{equation}
    H_Q=-\tau_yH_Q^T\tau_y,
\end{equation}
which implies $A=0$. The anticommuting chiral symmetry requires
\begin{equation}
    H_Q=-\tau_zH_Q\tau_z,
\end{equation}
implying $B=0$. $C_{2z}T$ requires
\begin{equation}
    H_Q=\tau_xH_Q^*\tau_x,
\end{equation}
implying $B$ is imaginary whereas $C$ and $D$ are real.

In generality, $C_{2x}T$ requires
\begin{equation}
    H_Q=-\tau_zH_{M_yQ}^*\tau_z,
\end{equation}
whereas $C_{2y}$ requires
\begin{equation}
    H_Q=-\tau_yH_{M_xQ}\tau_y.
\end{equation}
However, the specific constraints of this action on the coefficients $A$-$D$ depend on the particular $\hat Q$ involved. When the set of $C_{3z}$-related $Q$s are closed under $M_x=-M_y$, these properties more stringently constrain the coefficients $A$-$D$. This is for instance the case for the first and third harmonics, where one of the $Q$ is in the $\pm\hat y$ direction, so the $C_{3z}$-related set of $Q$s are closed under $M_x=-M_y$. In these harmonics, $C_{2x}$ implies that $B$ and $C$ are imaginary, whereas $D$ is real; while $C_{2y}$ implies that $C=0$. 

However, not all harmonics have a $Q$ in the $\pm\hat y$ direction; for these higher harmonics, $C_{2x}$ and $C_{2y}$ impose the constraint that the coefficients of a $C_{3z}$-related set of $Q$s is related to its corresponding mirror.

\subsection{Nonzero $K$-to-$K$ harmonics: $H_{G,\pm}$}

Time-reversal symmetry implies
\begin{equation}
    H_{G,\pm}=\tau_yH_{-G,\mp}^*\tau_y,
\end{equation}
which together with hermiticity, $H_{-G}=H_G^\dagger$, imposes the constraint
\begin{equation}\label{eq_hhTsym}
H_{G,\pm}=\tau_yH_{G,\mp}^T\tau_y.
\end{equation}
Therefore, $A$ is SV-even, whereas all other coefficients are SV-odd. Chiral symmetry requires
\begin{equation}
    H_{G,\pm}=-\tau_zH_{G,\pm}\tau_z, 
\end{equation}
implying $A=B=0$. $C_{2z}T$ requires
\begin{equation}
    H_{G,\pm}=\tau_xH_{G,\pm}^*\tau_x,
\end{equation}
implying that $A$, $C$, and $D$ are real, whereas $B$ is imaginary.

Like $H_Q$, for $H_{G,\pm}$, specific constraints of $C_{2x}$ and $C_{2y}$ on the coefficients $A$-$D$ depend on the specific $G$. In generality, $C_{2x}T$ requires
\begin{equation}
    H_{G,\pm}=\tau_zH_{M_yG,\pm}^T\tau_z,
\end{equation}
whereas $C_{2y}$ requires
\begin{equation}
    H_{G,\pm}=\tau_yH_{M_xG,\pm}\tau_y.
\end{equation}

There are three possibilities for the symmetry of a $C_{3z}$+$T$-related set of $G$s under reflections $R_x$ and $R_y$. They may not be closed under these operations, in which case coefficients at $G$ and $M_xG$ are related. If the set of $G$ for some harmonic are closed under $M_x$ and $M_y$, then there will be some $G$ in the set which is left fixed under either $M_x$ or $M_y$; the constraints on the coefficients depend on which reflection leaves a $G$ fixed.

For the second harmonic, one $G$ lies along the $\hat x$ direction, so $M_y$ preserves that $G$. For this case, $C_{2x}$ implies that $C=0$, whereas $C_{2y}$ implies that $A$ and $C$ are real while $B$ and $D$ are imaginary.

For the alternative possible higher harmonic, where $M_x$ leaves a $G$ fixed, we find different constraints. $C_{2x}$ implies that $A$, $B$, and $C$ are real, whereas $D$ is imaginary. $C_{2y}$ implies that $B=D=0$.

\section{Additional candidate substrates \label{Apx:mat_tables}}

See Tables~\ref{tab:other_stable} and~\ref{tab:metastable}.

\onecolumngrid

\begin{table}[]
\centering
\begin{tabular}{|lll|lcc|ccc|c|}
\hline
 & Composition & mp-id & Surface & Lattice & SG & $|\epsilon|$\% & $V_1^*$ (meV) & \Mr length (nm) & Gap (eV) \\
\hline
\hline
\cline{1-10}
 & NaNdSe$_2$ & \href{https://materialsproject.org/materials/mp-999471?material_ids=mp-999471}{999471} & 111 & rhomb. & $R\bar{3}m$ & 0.13 & 9 & 1905 & 1.83 \\
\cline{1-10}
 & RbH & \href{https://materialsproject.org/materials/mp-24721?material_ids=mp-24721}{24721} & 111 & cubic & $Fm\bar{3}m$ & 0.14 & 9 & 1777 & 3.20 \\
\cline{1-10}
 & CsPrS$_2$ & \href{https://materialsproject.org/materials/mp-9037?material_ids=mp-9037}{9037} & 001 & hex. & $P6_3/mmc$ & 0.19 & 13 & 1272 & 2.28 \\
\cline{1-10}
 & RbTbSe$_2$ & \href{https://materialsproject.org/materials/mp-10782?material_ids=mp-10782}{10782} & 111 & rhomb. & $R\bar{3}m$ & 0.27 & 18 & 907 & 2.01 \\
\cline{1-10}
 & SmTlSe$_2$ & \href{https://materialsproject.org/materials/mp-999137?material_ids=mp-999137}{999137} & 111 & rhomb. & $R\bar{3}m$ & 0.28 & 18 & 897 & 1.42 \\
\cline{1-10}
 & NaBr & \href{https://materialsproject.org/materials/mp-22916?material_ids=mp-22916}{22916} & 111 & cubic & $Fm\bar{3}m$ & 0.29 & 19 & 841 & 4.09 \\
\cline{1-10}
 & SrS & \href{https://materialsproject.org/materials/mp-1087?material_ids=mp-1087}{1087} & 111 & cubic & $Fm\bar{3}m$ & 0.30 & 20 & 828 & 2.56 \\
\cline{1-10}
 & RbPrS$_2$ & \href{https://materialsproject.org/materials/mp-9362?material_ids=mp-9362}{9362} & 111 & rhomb. & $R\bar{3}m$ & 0.32 & 21 & 774 & 2.36 \\
\cline{1-10}
 & Li$_2$Se & \href{https://materialsproject.org/materials/mp-2286?material_ids=mp-2286}{2286} & 111 & cubic & $Fm\bar{3}m$ & 0.42 & 28 & 586 & 3.11 \\
\cline{1-10}
 & Sr(Li$_2$P)$_2$ & \href{https://materialsproject.org/materials/mp-570097?material_ids=mp-570097}{570097} & 111 & rhomb. & $R\bar{3}m$ & 0.47 & 31 & 522 & 1.17 \\
\cline{1-10}
 & CsNdS$_2$ & \href{https://materialsproject.org/materials/mp-1095172?material_ids=mp-1095172}{1095172} & 001 & hex. & $P6_3/mmc$ & 0.55 & 36 & 450 & 2.27 \\
\cline{1-10}
 & KLaS$_2$ & \href{https://materialsproject.org/materials/mp-15781?material_ids=mp-15781}{15781} & 111 & rhomb. & $R\bar{3}m$ & 0.57 & 37 & 434 & 2.77 \\
\cline{1-10}
 & BiTeCl & \href{https://materialsproject.org/materials/mp-28944?material_ids=mp-28944}{28944} & 001 & hex. & $P6_3mc$ & 0.66 & 43 & 375 & 1.56 \\
\cline{1-10}
 & U(OF)$_2$ & \href{https://materialsproject.org/materials/mp-27980?material_ids=mp-27980}{27980} & 111 & rhomb. & $R\bar{3}m$ & 0.67 & 44 & 367 & 2.12 \\
\cline{1-10}
 & MgSe & \href{https://materialsproject.org/materials/mp-13031?material_ids=mp-13031}{13031} & 111 & cubic & $F\bar{4}3m$ & 0.73 & 48 & 336 & 2.55 \\
\cline{1-10}
 & GeI$_2$ & \href{https://materialsproject.org/materials/mp-27922?material_ids=mp-27922}{27922} & 001 & hex. & $P\bar{3}m1$ & 0.76 & 50 & 325 & 2.07 \\
\cline{1-10}
 & KSmSe$_2$ & \href{https://materialsproject.org/materials/mp-1006891?material_ids=mp-1006891}{1006891} & 111 & rhomb. & $R\bar{3}m$ & 0.77 & 51 & 320 & 1.97 \\
\cline{1-10}
 & PbF$_2$ & \href{https://materialsproject.org/materials/mp-315?material_ids=mp-315}{315} & 111 & cubic & $Fm\bar{3}m$ & 0.79 & 52 & 314 & 4.44 \\
\cline{1-10}
 & NaPrSe$_2$ & \href{https://materialsproject.org/materials/mp-999461?material_ids=mp-999461}{999461} & 111 & rhomb. & $R\bar{3}m$ & 0.80 & 53 & 307 & 1.83 \\
\cline{1-10}
 & Ca$_2$PI & \href{https://materialsproject.org/materials/mp-23040?material_ids=mp-23040}{23040} & 111 & rhomb. & $R\bar{3}m$ & 0.83 & 55 & 297 & 1.80 \\
\cline{1-10}
 & LaSeF & \href{https://materialsproject.org/materials/mp-7738?material_ids=mp-7738}{7738} & 001 & hex. & $P6_3/mmc$ & 0.90 & 59 & 275 & 1.72 \\
\cline{1-10}
 & KPrS$_2$ & \href{https://materialsproject.org/materials/mp-15782?material_ids=mp-15782}{15782} & 111 & rhomb. & $R\bar{3}m$ & 0.95 & 62 & 259 & 2.36 \\
\cline{1-10}
 & RbHoSe$_2$ & \href{https://materialsproject.org/materials/mp-10783?material_ids=mp-10783}{10783} & 111 & rhomb. & $R\bar{3}m$ & 1.07 & 70 & 231 & 2.04 \\
\cline{1-10}
 & RbNdS$_2$ & \href{https://materialsproject.org/materials/mp-9363?material_ids=mp-9363}{9363} & 111 & rhomb. & $R\bar{3}m$ & 1.07 & 71 & 230 & 2.35 \\
\cline{1-10}
 & CsF & \href{https://materialsproject.org/materials/mp-1784?material_ids=mp-1784}{1784} & 111 & cubic & $Fm\bar{3}m$ & 1.19 & 78 & 207 & 5.28 \\
\cline{1-10}
 & RbLaS$_2$ & \href{https://materialsproject.org/materials/mp-9361?material_ids=mp-9361}{9361} & 111 & rhomb. & $R\bar{3}m$ & 1.20 & 79 & 206 & 2.75 \\
\cline{1-10}
 & Ba$_2$CuClO$_2$ & \href{https://materialsproject.org/materials/mp-551456?material_ids=mp-551456}{551456} & 111 & rhomb. & $R\bar{3}m$ & 1.29 & 85 & 191 & 2.31 \\
\cline{1-10}
 & RbSmSe$_2$ & \href{https://materialsproject.org/materials/mp-10780?material_ids=mp-10780}{10780} & 111 & rhomb. & $R\bar{3}m$ & 1.30 & 86 & 189 & 1.98 \\
\cline{1-10}
 & CaSe & \href{https://materialsproject.org/materials/mp-1415?material_ids=mp-1415}{1415} & 111 & cubic & $Fm\bar{3}m$ & 1.31 & 86 & 188 & 2.14 \\
\cline{1-10}
 & NdTlSe$_2$ & \href{https://materialsproject.org/materials/mp-568588?material_ids=mp-568588}{568588} & 111 & rhomb. & $R\bar{3}m$ & 1.31 & 86 & 188 & 1.44 \\
\cline{1-10}
 & RbErSe$_2$ & \href{https://materialsproject.org/materials/mp-10784?material_ids=mp-10784}{10784} & 111 & rhomb. & $R\bar{3}m$ & 1.41 & 93 & 175 & 2.06 \\
\cline{1-10}
 & KNdS$_2$ & \href{https://materialsproject.org/materials/mp-1006885?material_ids=mp-1006885}{1006885} & 111 & rhomb. & $R\bar{3}m$ & 1.44 & 95 & 171 & 2.32 \\
\cline{1-10}
 & MgI$_2$ & \href{https://materialsproject.org/materials/mp-23205?material_ids=mp-23205}{23205} & 001 & hex. & $P\bar{3}m1$ & 1.49 & 98 & 165 & 3.68 \\
\cline{1-10}
 & RbNdSe$_2$ & \href{https://materialsproject.org/materials/mp-10779?material_ids=mp-10779}{10779} & 111 & rhomb. & $R\bar{3}m$ & 2.52 & 165 & 98 & 1.97 \\
\cline{1-10}
 & SrLiP & \href{https://materialsproject.org/materials/mp-13276?material_ids=mp-13276}{13276} & 001 & hex. & $P6_3/mmc$ & 2.54 & 167 & 97 & 1.35 \\
\cline{1-10}
 & NaLaSe$_2$ & \href{https://materialsproject.org/materials/mp-999472?material_ids=mp-999472}{999472} & 111 & rhomb. & $R\bar{3}m$ & 2.55 & 168 & 97 & 2.35 \\
\cline{1-10}
 & CsGdS$_2$ & \href{https://materialsproject.org/materials/mp-9084?material_ids=mp-9084}{9084} & 111 & rhomb. & $R\bar{3}m$ & 2.56 & 168 & 96 & 1.77 \\
\cline{1-10}
 & ErTlSe$_2$ & \href{https://materialsproject.org/materials/mp-570117?material_ids=mp-570117}{570117} & 111 & rhomb. & $R\bar{3}m$ & 2.58 & 169 & 96 & 1.40 \\
\cline{1-10}
 & Ba$_2$NCl & \href{https://materialsproject.org/materials/mp-1018099?material_ids=mp-1018099}{1018099} & 111 & rhomb. & $R\bar{3}m$ & 2.64 & 174 & 93 & 1.28 \\
\cline{1-10}
 & Ba(MgP)$_2$ & \href{https://materialsproject.org/materials/mp-8278?material_ids=mp-8278}{8278} & 001 & hex. & $P\bar{3}m1$ & 2.67 & 176 & 92 & 1.13 \\
\cline{1-10}
 & NaDySe$_2$ & \href{https://materialsproject.org/materials/mp-999488?material_ids=mp-999488}{999488} & 111 & rhomb. & $R\bar{3}m$ & 2.73 & 180 & 90 & 1.92 \\
\cline{1-10}
 & Sr$_2$BBrN$_2$ & \href{https://materialsproject.org/materials/mp-1077553?material_ids=mp-1077553}{1077553} & 111 & rhomb. & $R\bar{3}m$ & 2.81 & 185 & 88 & 2.45 \\
\cline{1-10}
 & SrF$_2$ & \href{https://materialsproject.org/materials/mp-981?material_ids=mp-981}{981} & 111 & cubic & $Fm\bar{3}m$ & 2.84 & 187 & 87 & 6.78 \\
\cline{1-10}
\end{tabular}
\caption{\label{tab:other_stable}
  Additional thermodynamically stable systems for a wider range of $V_1^*$ than reported in Table~\ref{tab:Materials} in the main text.
}
\end{table}

\begin{table}[]
\centering
\begin{tabular}{|lll|lcc|ccc|cc|}
\hline
  & Composition & mp-id & Surface & Lattice & SG & $|\epsilon|$\% & $V_1^*$ (meV) & \Mr length (nm) & Gap (eV) & $E_\text{hull}$ (meV) \\
\hline
\hline
 & CuI & \href{https://materialsproject.org/materials/mp-22895?material_ids=mp-22895}{22895} & 111 & cubic & $F\bar{4}3m$ & 0.16 & 11 & 150.13 & 1.18 & 6.76 \\
\cline{1-11}
 & NaBH$_4$ & \href{https://materialsproject.org/materials/mp-976181?material_ids=mp-976181}{976181} & 111 & cubic & $F\bar{4}3m$ & 0.29 & 19 & 86.15 & 6.59 & 11.45 \\
\cline{1-11}
 & CsPrS$_2$ & \href{https://materialsproject.org/materials/mp-9080?material_ids=mp-9080}{9080} & 111 & rhomb. & $R\bar{3}m$ & 0.30 & 20 & 81.98 & 2.26 & 1.41 \\
\cline{1-11}
 & LiI & \href{https://materialsproject.org/materials/mp-22899?material_ids=mp-22899}{22899} & 111 & cubic & $Fm\bar{3}m$ & 0.32 & 21 & 76.05 & 4.24 & 36.53 \\
\cline{1-11}
 & CuI & \href{https://materialsproject.org/materials/mp-569346?material_ids=mp-569346}{569346} & 001 & hex. & $P6_3mc$ & 0.37 & 25 & 66.01 & 1.21 & 9.15 \\
\cline{1-11}
 & CsNdS$_2$ & \href{https://materialsproject.org/materials/mp-9081?material_ids=mp-9081}{9081} & 111 & rhomb. & $R\bar{3}m$ & 0.46 & 30 & 54.02 & 2.25 & 1.53 \\
\cline{1-11}
 & MgSe & \href{https://materialsproject.org/materials/mp-1018040?material_ids=mp-1018040}{1018040} & 001 & hex. & $P6_3mc$ & 0.49 & 32 & 50.74 & 2.58 & 42.17 \\
\cline{1-11}
 & Sr$_2$H$_3$I & \href{https://materialsproject.org/materials/mp-1019269?material_ids=mp-1019269}{1019269} & 001 & hex. & $P\bar{3}m1$ & 0.61 & 40 & 40.17 & 2.15 & 8.07 \\
\cline{1-11}
 & LiI & \href{https://materialsproject.org/materials/mp-568273?material_ids=mp-568273}{568273} & 001 & hex. & $P6_3/mmc$ & 0.69 & 45 & 35.93 & 4.24 & 33.21 \\
\cline{1-11}
 & PbS & \href{https://materialsproject.org/materials/mp-1057015?material_ids=mp-1057015}{1057015} & 111 & rhomb. & $R3m$ & 0.70 & 46 & 35.47 & 1.04 & 1.34 \\
\cline{1-11}
 & CdI$_2$ & \href{https://materialsproject.org/materials/mp-570437?material_ids=mp-570437}{570437} & 001 & hex. & $P3m1$ & 1.30 & 86 & 18.95 & 2.40 & 1.25 \\
\cline{1-11}
 & CdI$_2$ & \href{https://materialsproject.org/materials/mp-568289?material_ids=mp-568289}{568289} & 111 & rhomb. & $R\bar{3}m$ & 1.34 & 88 & 18.47 & 2.38 & 8.45 \\
\cline{1-11}
 & CdI$_2$ & \href{https://materialsproject.org/materials/mp-567296?material_ids=mp-567296}{567296} & 001 & hex. & $P\bar{3}m1$ & 1.34 & 88 & 18.44 & 2.38 & 1.52 \\
\cline{1-11}
 & KLiS & \href{https://materialsproject.org/materials/mp-1077266?material_ids=mp-1077266}{1077266} & 001 & hex. & $P6_3/mmc$ & 1.35 & 89 & 18.30 & 2.74 & 48.94 \\
\cline{1-11}
 & CdI$_2$ & \href{https://materialsproject.org/materials/mp-28248?material_ids=mp-28248}{28248} & 001 & hex. & $P6_3mc$ & 1.40 & 92 & 17.67 & 2.44 & 0.62 \\
\cline{1-11}
 & CdI$_2$ & \href{https://materialsproject.org/materials/mp-567259?material_ids=mp-567259}{567259} & 001 & hex. & $P\bar{3}m1$ & 1.67 & 110 & 14.75 & 2.40 & 1.79 \\
\cline{1-11}
 & CdS & \href{https://materialsproject.org/materials/mp-2469?material_ids=mp-2469}{2469} & 111 & cubic & $F\bar{4}3m$ & 1.69 & 111 & 14.64 & 1.05 & 1.46 \\
\cline{1-11}
 & ScP & \href{https://materialsproject.org/materials/mp-1009746?material_ids=mp-1009746}{1009746} & 111 & cubic & $F\bar{4}3m$ & 1.79 & 117 & 13.82 & 1.68 & 530.03 \\
\cline{1-11}
 & LaF$_3$ & \href{https://materialsproject.org/materials/mp-8354?material_ids=mp-8354}{8354} & 001 & hex. & $P6_3/mmc$ & 1.86 & 122 & 13.24 & 5.74 & 19.58 \\
\cline{1-11}
 & AgI & \href{https://materialsproject.org/materials/mp-22919?material_ids=mp-22919}{22919} & 111 & cubic & $Fm\bar{3}m$ & 2.02 & 133 & 12.20 & 1.01 & 92.18 \\
\cline{1-11}
 & ZnTe & \href{https://materialsproject.org/materials/mp-8884?material_ids=mp-8884}{8884} & 001 & hex. & $P6_3mc$ & 2.10 & 138 & 11.75 & 1.10 & 5.24 \\
\cline{1-11}
 & NaH & \href{https://materialsproject.org/materials/mp-1009220?material_ids=mp-1009220}{1009220} & 111 & cubic & $Pm\bar{3}m$ & 2.16 & 142 & 11.41 & 1.50 & 157.43 \\
\cline{1-11}
 & ZnTe$^*$ & \href{https://materialsproject.org/materials/mp-571195?material_ids=mp-571195}{571195} & 001 & hex. & $P3_1$ & 2.26 & 149 & 10.90 & 1.07 & 3.69 \\
\cline{1-11}
 & GeI$_2$ & \href{https://materialsproject.org/materials/mp-567677?material_ids=mp-567677}{567677} & 001 & hex. & $P\bar{6}m2$ & 2.55 & 168 & 9.68 & 2.17 & 150.90 \\
\cline{1-11}
 & SrF$_2$ & \href{https://materialsproject.org/materials/mp-1019258?material_ids=mp-1019258}{1019258} & 001 & hex. & $P6_3/mmc$ & 2.76 & 181 & 8.94 & 5.94 & 170.96 \\
\cline{1-11}
\end{tabular}
\caption{ \label{tab:metastable}
  Candidate substrates with formation energies above the convex hull of the Gibbs free energy. While in principle they can be thermodynamically unstable, they may form long lived metastable states. Formation energies above the convex hull are given in the far right column. The $^*$ on ZnTe in the $P3_1$ phase indicates that this is the only substrate without mirror symmetry. 
}
\end{table}

\end{document}